 \documentclass[final,5p,times,twocolumn]{elsarticle}

\usepackage{amssymb,amsthm,amssymb,mathrsfs,amsmath}
\usepackage{color}

\newtheorem{thm}{Theorem}[section]
\newtheorem{prop}{Proposition}[section]

\newtheorem{rmk}{\it Remark}[section]

\newcommand{\I}{\mathrm{Im}}

\newcommand{\eps}{\epsilon}
\newcommand{\bo}{\mathcal{O}}

\def\Xint#1{\mathchoice
{\XXint\displaystyle\textstyle{#1}}%
{\XXint\textstyle\scriptstyle{#1}}%
{\XXint\scriptstyle\scriptscriptstyle{#1}}%
{\XXint\scriptscriptstyle\scriptscriptstyle{#1}}%
\!\int}
\def\XXint#1#2#3{{\setbox0=\hbox{$#1{#2#3}{\int}$ }
\vcenter{\hbox{$#2#3$ }}\kern-.6\wd0}}

\def\dashint{\Xint-}

\journal{Physica D}

\begin{document}

\begin{frontmatter}



\title{On the generation of dispersive shock waves\tnoteref{toWhitham}}
\tnotetext[toWhitham]{In memory of G. B. Whitham.}


\author{Peter D. Miller}\ead{millerpd@umich.edu}
\address{Department of Mathematics, University of Michigan, East Hall, 530 Church St., Ann Arbor, MI 48109}

\begin{abstract}
We review various methods for the analysis of initial-value problems for integrable dispersive equations in the weak-dispersion or semiclassical regime.  Some methods are sufficiently powerful to rigorously explain the generation of modulated wavetrains, so-called dispersive shock waves, as the result of shock formation in a limiting dispersionless system.  They also provide a detailed description of the solution near caustic curves that delimit dispersive shock waves, revealing fascinating universal wave patterns. 
\end{abstract}

\begin{keyword}
Semiclassical limit \sep small-dispersion limit \sep Lax-Levermore theory \sep Deift-Zhou steepest descent method \sep universality.
\end{keyword}

\end{frontmatter}

\section{Introduction}\label{introduc}
Many physical systems exhibit behavior that can be approximated by periodic traveling wave solutions of partial differential equations (PDEs).  G.~B.~Whitham realized that the particular periodic wave that best fits the observed wave field is frequently different depending upon where and when the observation is made, leading to the notion of a slowly-modulated wavetrain.  There is a rather complete theory of modulated waves developed originally by Whitham with subsequent contributions by many others, some of which are described in other papers of this volume.  

The main hypothesis of the theory is the existence of an approximate solution of the equation at hand that has the form of a periodic wave whose parameters (e.g., amplitude, wavenumber, etc.) depend slowly on space and time.  The relative slowness of the modulation relative to the wavelength and period of the wave is mathematically built in via an artificial small parameter $\epsilon$ measuring the ratio between the microscopic (wavelength and period) and macroscopic (modulation) scales.  By various asymptotic methods all ultimately employing some kind of averaging over the microstructure, the limit $\eps\to 0$ allows one to deduce a closed system of PDEs governing the slowly-varying parameters alone, the famous \emph{Whitham modulation equations}.

A valid question is whether Whitham's theory only describes the evolution in time of a pre-existing modulated wave (i.e., an initial condition having this form), or if not, how such structures are generated automatically by the underlying system.  To properly formulate this question, we choose initial data having no fast oscillations at all, and ask whether the oscillations appear on their own after some time, modeling the formation of a \emph{dispersive shock wave} (DSW).  Mathematically, it is convenient to scale the spatial independent variables so that the (slowly-varying) initial data is fixed in the limit $\eps\to 0$, and anticipating an initial slow phase of the dynamics, time should be similarly rescaled.  This means that the exact periodic wave solutions should have wavelengths and periods proportional to $\eps$, and thus the rule for rescaling a given dispersive equation is simply to introduce the small parameter by the replacements $\partial_x\mapsto \eps\partial_x$ and $\partial_t\mapsto \eps\partial_t$.  For example, if one applies this rule to the Korteweg-de Vries (KdV) equation $u_t + 2uu_x +\tfrac{1}{3}u_{xxx}=0$, after canceling a factor of $\eps\neq 0$ one arrives at the form
\begin{equation}
u_t + 2uu_x +\tfrac{1}{3}\eps^2u_{xxx}=0
\label{eq:KdV}
\end{equation}
which should be formulated with initial data $u(x,0)=u_0(x)$ independent of $\eps$.  In this case, the limit $\eps\to 0$ that separates the scales in the Whitham theory can be interpreted as the \emph{small-dispersion limit}.

As another example, we may apply the rescaling rule to the nonlinear Schr\"odinger (NLS) equation $i\psi_t +\tfrac{1}{2}\psi_{xx}-\sigma|\psi|^2\psi=0$ (with parameter $\sigma\in\mathbb{R}$) which becomes
\begin{equation}
i\eps\psi_t +\tfrac{1}{2}\eps^2\psi_{xx}-\sigma|\psi|^2\psi=0.
\label{eq:NLS-general}
\end{equation}
For physical reasons it is frequently important to consider this equation with initial data that includes $\epsilon$-dependence in the form of a fast phase:  $\psi(x,0)=A(x)e^{iS(x)/\eps}$, where the positive amplitude $A$ and real phase $S$ are independent of $\eps$.  In this case, due to the fact that in free-particle quantum mechanics ($\sigma=0$) the parameter $\eps>0$ is proportional to Planck's constant, the limit $\eps\to 0$ in this context is frequently called the \emph{semiclassical limit}.  It may seem to be strange terminology, since setting $\epsilon=0$ to be ``in the limit'' leaves an equation that is trivial and the initial data makes no sense unless $S\equiv 0$.  However, some sense is restored upon introducing the ``density'' and ``velocity'' variables $\rho:=|\psi|^2$ and $u:=\eps\I\{\psi_x/\psi\}$, respectively, for a ``quantum fluid''.  Indeed, Madelung \cite{Madelung26} showed that the 
equation \eqref{eq:NLS-general} implies the following closed system:
\begin{equation}
\rho_t + (\rho u)_x=0\quad \text{and}\quad
u_t+\left(\tfrac{1}{2}u^2+\sigma\rho\right)_x=\tfrac{1}{2}\eps^2F[\rho]_x,
\label{eq:Madelung-general}
\end{equation}
where
\begin{equation}
F[\rho]:=\frac{\rho_{xx}}{2\rho}-\left(\frac{\rho_x}{2\rho}\right)^2.
\end{equation}
For the initial data $\psi(x,0)=\psi_0(x)=A(x)e^{iS(x)/\eps}$, one has corresponding $\eps$-independent initial data $\rho(x,0)=\rho_0(x)=A(x)^2$ and $u(x,0)=u_0(x)=S'(x)$.  It now appears attractive to define the limiting dynamics by simply setting $\eps=0$ in \eqref{eq:Madelung-general}, resulting in the \emph{dispersionless NLS system}.  This system is hyperbolic for $\sigma>0$ (defocusing case), elliptic for $\sigma<0$ (focusing case), and degenerate hyperbolic for $\sigma=0$ (linear case).

This paper is a survey of techniques available for the analysis of small-dispersion or semiclassical limits for dispersive wave equations, and includes as well some results that have been established by their means.  The aim of these techniques is, in general, to explain as much as possible the spontaneous onset of modulated oscillations from relatively smooth initial data.


\section{General asymptotic methods for initial-value problems}
\subsection{Matching modulated wavetrains and dispersionless fields}
\label{sec:GP}
In 1973, Gurevich and Pitaevskii \cite{GurevichP73} designed a method for explaining the generation of modulated waves in the context of the initial-value problem for the weakly-dispersive KdV equation \eqref{eq:KdV}.  Their method, which applies equally well to some non-integrable equations, was essentially to match a solution of Whitham's modulation equations for periodic traveling waves occupying an interval $x^-(t)<x<x^+(t)$ onto solutions of the inviscid Burgers (IB) equation $u_t+2uu_x=0$ (the dispersionless limit) for $x<x^-(t)$ and $x>x^+(t)$.  In the process, they determined the functions $x^\pm(t)$ bounding the DSW in the $(x,t)$-plane.  The method described in \cite{GurevichP73} works because when written in Riemann-invariant form, Whitham's equations for three fields $u_1<u_2<u_3$ degenerate to the IB equation for $u_1$ when $u_2=u_3$, and for $u_3$ when $u_1=u_2$.  It follows that the method of \cite{GurevichP73} may be described as seeking a global weak solution of the Whitham modulation equations that exhibits certain degeneracies for small $t$ and large $x$.  This type of degeneration is a general and expected property of Whitham modulation equations, so many problems have been studied in this way; it was used, for example, in \cite{BlochK92} to study the Toda lattice.  It should be observed, however, that while very reasonable and delivering of physically satisfying results, this method is not one of rigorous analysis because there is no obvious way to estimate the errors in this approach or otherwise to prove its convergence as $\eps\to 0$.

\subsection{Rigorous pre-breaking analysis}
It is possible to obtain quite general and far-reaching convergence results for weakly-dispersive or semiclassical initial-value problems, provided one is interested in the dynamics that occur before singularities appear in the solution of the approximating ($\eps=0$) limiting equation/system for the same initial data.  While the details are complicated, the basic idea is one of regular perturbation theory, in which the effects of terms in \eqref{eq:KdV} or \eqref{eq:Madelung-general} proportional to powers of $\eps$ are controlled by functional analytic estimations relying on the smoothness of the approximating solutions.  Grenier \cite{Grenier95} used such an approach to establish the semiclassical limit for quite general NLS equations with defocusing nonlinearities.  For the more challenging focusing type of nonlinearity, G\'erard \cite{Gerard93} obtained analogous results with the additional assumption of analyticity of the initial data, which allows for the solution of the approximating quasilinear system (the analogue of \eqref{eq:Madelung-general} with $\sigma=-1$ and $\eps=0$) by convergent Cauchy-Kovaleskaya series.  While these rigorous convergence results are valid for NLS equations in more than one space dimension and with general nonlinearities, they fail as soon as the problem becomes interesting with the formation of a singularity that one expects dispersion to regularize via the generation of a DSW.

\section{Transform-based global analysis for integrable PDEs}
When we turn our attention to the problem of global analysis for initial-value problems of weakly dispersive waves, i.e., the determination of properties of the solution for times larger than the breaking time for the dispersionless approximation, most rigorous results have been obtained for problems that are integrable by means of a \emph{scattering transform} (ST).  A ST provides a roadmap for the construction of the solution of the initial-value problem and is a nonlinear generalization of the Fourier transform pair.  Hand-in-hand with the special property of a dispersive equation being integrable by means of a ST come several remarkable structural properties of the periodic solutions and their Whitham modulation equations \cite{FlaschkaFM80}.  Indeed, integrable problems admit not just periodic traveling wave solutions, but also rich families of multiphase waves (nonlinear superpositions of several co- or counter-propagating periodic waves) that have a natural algebro-geometric characterization in terms of the function theory of certain Riemann surfaces.  Such multiphase waves can be averaged to obtain a macroscopic description of their modulated dynamics by means of multiphase Whitham equations.  It was shown in \cite{FlaschkaFM80} that it is a general property of integrable systems that the multiphase Whitham equations can be cast into Riemann-invariant form, a property that is generally not held by quasilinear systems of more than two unknowns.  The structure of such nearly diagonal systems arising from integrable wave equations is sufficiently special that it is possible to obtain general solutions of these systems by a generalization of the hodograph method found by Tsar\"ev \cite{Tsarev85}.  In some way, all of these properties play a role in the rigorous global analysis of weakly dispersive initial-value problems for integrable equations.

Another general aspect of the use of a ST to study problems of weak dispersion is that the parameter $\eps$ enters into the problem of computing the scattering data.  This is both a blessing and a curse; the good news is that it makes available approximations based on the WKB method to calculate the direct ST, but the bad news is that qualitatively similar initial conditions can generate scattering data of such different types that substantially different methods of analysis are required to analyze the inverse ST.  For example, it is a familiar fact that the scattering data for the ST solution of the initial-value problem for the KdV equation \eqref{eq:KdV} with decaying data $u_0$ splits into a contribution from a reflection coefficient (continuous spectrum) and contributions from a number of eigenvalues (discrete spectrum).  It turns out that if $u_0$ is a negative bell-shaped function then there are no eigenvalues and the whole solution is generated from the reflection coefficient that can be calculated in the limit $\eps\to 0$; if on the other hand $u_0$ is a positive bell-shaped function then in the same limit the reflection coefficient vanishes while a large ($\sim \eps^{-1}$) number of eigenvalues are produced.
This means that different methods are required for the analysis of the inverse problem in these two cases.  One unfortunate implication of this phenomenon is that for any given equation there are two or more different ``schools'' of authors writing papers on the topic based on their expertise with different techniques of analysis for the inverse problem.  Most of my own work has been on problems that produce a large discrete spectrum (see \cite{Miller08} for a review), but for the purposes of this article I have tried to be more even-handed.  

\subsection{The semiclassical linear Schr\"odinger equation}
\label{sec:linear}
To set the basic context, we begin with an elementary problem,
the free-particle (linear) Schr\"odinger equation, namely \eqref{eq:NLS-general} with $\sigma=0$,
for which we specify the initial data
\begin{equation}
\psi(x,0)=\psi_0(x)=
\sqrt{\rho_0(x)}e^{iS(x)/\epsilon},\; S(x):=\int_0^x u_0(y)\,dy.
\label{eq:schrod-initial-data}
\end{equation}
The overall goal is to describe how $\psi$ depends on 
the independent variables $x$ and $t$, 
the parameter
$\epsilon$, and 
the initial data $\psi_0$ (equivalently, the amplitude $\rho_0>0$ and phase gradient $u_0$).
The initial-value problem can be solved by the Fourier transform via the following familiar steps:
\begin{enumerate}
\item Direct transform:  $\displaystyle \hat{\psi}_0(\lambda):=\frac{1}{2\pi}\int_\mathbb{R}\psi_0(x)e^{2i\lambda x/\epsilon}\,dx$.
\item Time evolution:  $\hat{\psi}(\lambda,t)=e^{-2i\lambda^2t/\epsilon}\hat{\psi}_0(\lambda)$.
\item Inverse transform:  $\displaystyle \psi(x,t)=\frac{2}{\epsilon}\int_\mathbb{R}\hat{\psi}(\lambda,t)e^{-2i\lambda x/\epsilon}\,d\lambda$.
\end{enumerate}
Combining steps 2 and 3 gives an integral representation of $\psi(x,t)$ in terms of the transform $\hat{\psi}_0(\lambda)$:
\begin{equation}
\psi(x,t)=\frac{2}{\epsilon}\int_\mathbb{R}\hat{\psi}_0(\lambda)e^{-2i(\lambda x +\lambda^2 t)/\epsilon}\,d\lambda.
\end{equation}
Inserting the result of step 1 gives an iterated/double integral representation of $\psi(x,t)$ in terms of $\psi_0(x)$ directly:
\begin{equation}
\psi(x,t)=\frac{1}{\pi\epsilon}\int_\mathbb{R}\int_{\mathbb{R}}\psi_0(y)e^{-2i(\lambda(x-y) +\lambda^2 t)/\epsilon}\,dy\,d\lambda.
\label{eq:double-integral}
\end{equation}
One must regard as unusual those cases of initial data for which one may evaluate these integrals exactly.
Therefore, one turns to numerics or, as is our interest here, asymptotics.
A typical singular limit in which one can deduce information in general is the long-time limit $t\to\infty$.  Going into a moving frame with fixed velocity $v$ by writing $x=x_0+vt$, one may observe that since $x$ and $t$ only appear in the ``outer'' iterated integral it is enough to write
\begin{equation}
\psi(x_0+vt,t)=\frac{2}{\epsilon}\int_\mathbb{R}\hat{\psi}_0(\lambda)e^{-2i\lambda x_0/\epsilon}
e^{-2it(\lambda v+\lambda^2)/\epsilon}\,d\lambda.
\end{equation}
Applying the \emph{method of stationary phase} \cite[Chapter 5]{Miller06} in the case of one simple stationary phase point $\lambda=\lambda_\mathrm{c}:=-v/2=-(x-x_0)/(2t)$, one finds in the limit $t\to +\infty$
\begin{equation}
\psi(x_0+vt,t)=e^{-i\pi/4}\sqrt{\frac{2\pi}{\epsilon t}}\hat{\psi}_0(\lambda_\mathrm{c})e^{-2i\lambda_\mathrm{c}x_0/\epsilon}e^{2it\lambda_\mathrm{c}^2/\epsilon} + \bo(t^{-3/2}),
\end{equation}
a formula that gives rise to physical notions like group velocity and dispersion.  See \cite[Chapter 11]{Whitham99} for a particularly lucid account of this circle of ideas.  

\begin{figure}[h]
\begin{center}
\includegraphics[width=0.5\linewidth]{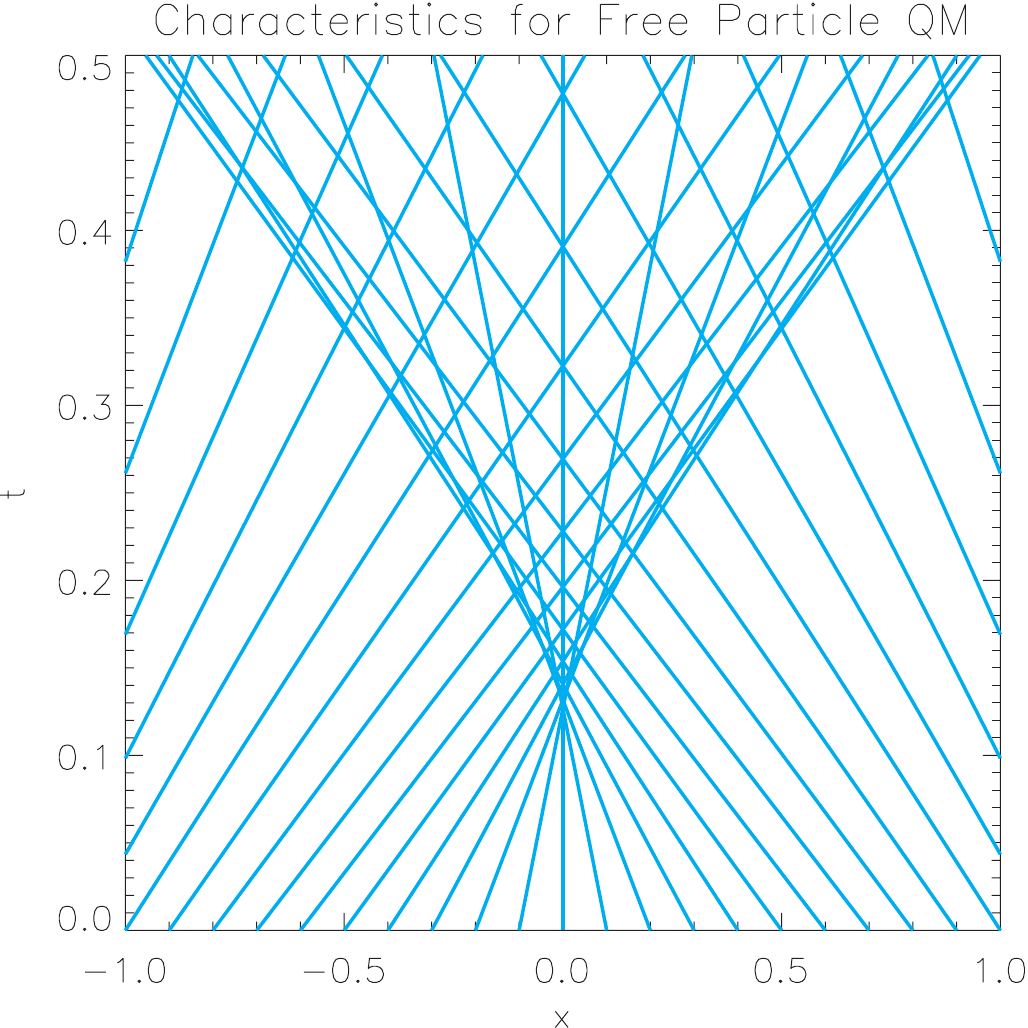}
\includegraphics[width=0.5\linewidth]{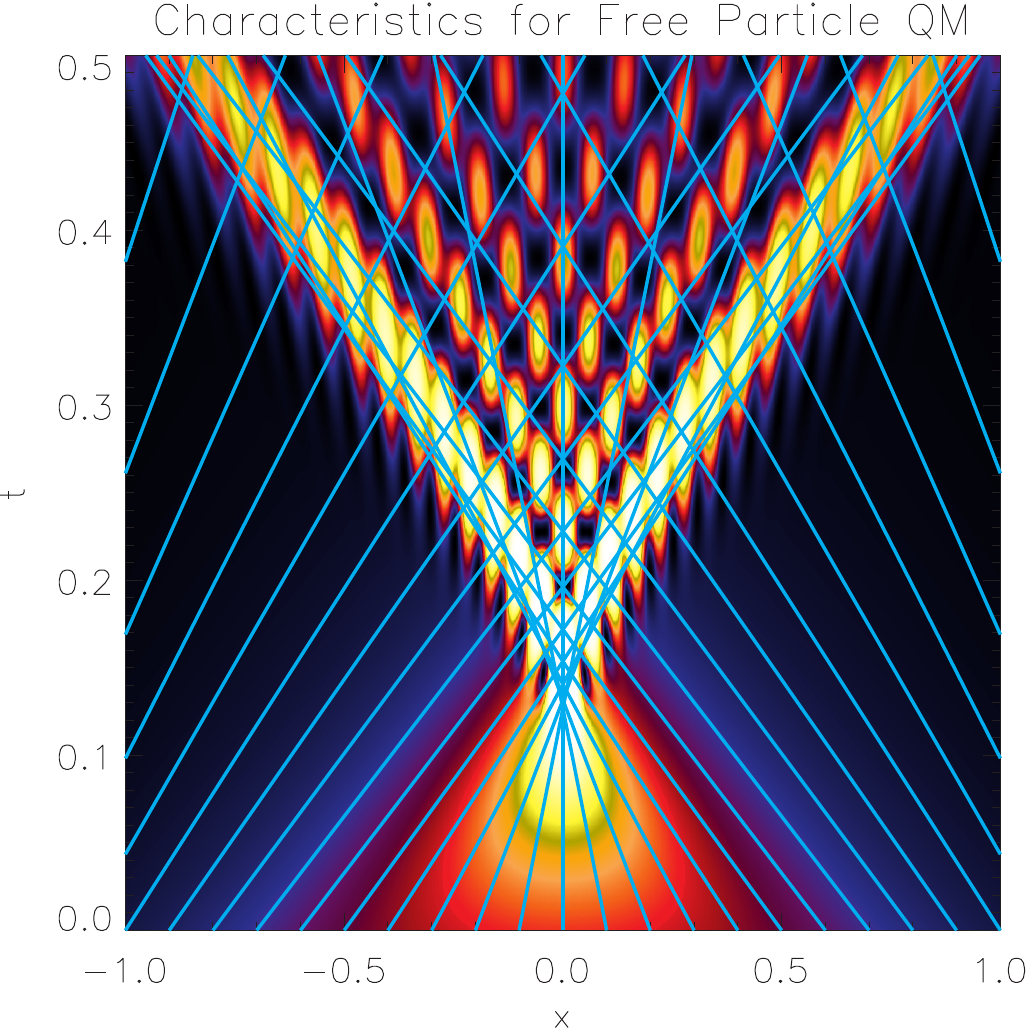}
\end{center}
\caption{Left:  the characteristic lines for the equation $u_t+uu_x=0$ for initial data $u_0(x)=-8\,\mathrm{sech}^2(x)\tanh(x)$.  Right:  the same superimposed on a density plot of $|\psi(x,t)|^2$ for $\eps=0.05$ for initial data consistent with the same $u_0$.}
\label{fig:SchrodingerCharacteristics}
\end{figure}

We may also consider the semiclassical limit $\epsilon\to 0$.  Now we must use the iterated/double integral formula \eqref{eq:double-integral}, but because \eqref{eq:NLS-general} with $\sigma=0$  has an exponential Green's function it is useful to carefully exchange the order of integration and reduce the problem again to a single integral: 
for $t>0$,
\begin{equation}
\psi(x,t)=\frac{e^{-i\pi/4}}{\sqrt{2\pi\epsilon t}}\int_{\mathbb{R}}e^{iI(y;x,t)/\epsilon}\sqrt{\rho_0(y)}\,dy,
\end{equation}
where the phase is $I(y;x,t):=S(y)+(y-x)^2/(2t)$.
Again the method of stationary phase applies, now to the limit $\epsilon\downarrow 0$:
\begin{equation}
\psi(x,t)=\frac{1}{\sqrt{t}}\sum_{n=0}^{2P}\frac{e^{i\pi((-1)^n-1)/4}}{\sqrt{|I''(y_n;x,t)|}}\sqrt{\rho_0(y_n)}e^{iI(y_n;x,t)/\epsilon} + \bo(\epsilon),
\label{eq:stationary-phase-formula}
\end{equation}
where $y_n=y_n(x,t)$, and $y_0<y_1<\cdots <y_{2P}$ are the stationary phase points, that is, the roots (assumed simple) of $I'(y;x,t)=0$.  
The condition that $y=y(x,t)$ is a stationary phase point is
$I'(y;x,t)=u_0(y)+(y-x)/t=0$, or equivalently for $t>0$, $x=u_0(y)t + y$.
The latter is exactly the equation for intercepts $y$ of characteristics through $(x,t)$ for the IB equation $u_t+uu_x=0$ arising from the formal limit ($\eps=0$) of the corresponding quantum hydrodynamic Madelung system \eqref{eq:Madelung-general} with $\sigma=0$. 
Fig.~\ref{fig:SchrodingerCharacteristics} makes clear the connection between the family of characteristic lines and the behavior of the solution $\psi(x,t)$ in various parts of the $(x,t)$-plane.
The formula \eqref{eq:stationary-phase-formula} shows that there is a critical time $t=t_c$ such that:
\begin{itemize}
\item If $t<t_c$ then there is just one characteristic line through each point and hence just one term in the sum.  Thus $\psi(x,t)$ looks like a modulated plane wave.
\item If $t>t_c$ then there are \emph{caustic curves} $x=x^\pm(t)$ with $x^\pm(t_c)=x_c$ such that
\begin{itemize}
\item If $x<x^-(t)$ or $x>x^+(t)$ then there is again just one characteristic through each point and again
$\psi(x,t)$ looks like a modulated plane wave.
\item If $x^-(t)<x<x^+(t)$ then there are \emph{three} lines through each point and hence three terms in the sum.  Interference among them makes $|\psi(x,t)|^2$ highly oscillatory.
\end{itemize}
\end{itemize}
The asymptotically abrupt transitions in the $(x,t)$-plane evident in Fig.~\ref{fig:SchrodingerCharacteristics} thus arise as bifurcation points for characteristics of the dispersionless problem, or equivalently, stationary phase points of an oscillatory integral.

\subsection{The semiclassical defocusing NLS equation}
Similar precision of analysis is available in principle for nonlinear dispersive wave problems that are integrable.
In place of the Fourier transform of the initial data (step 1), we have instead the \emph{direct ST}, which usually requires the analysis of a linear differential equation with a \emph{spectral parameter} $\lambda$ to obtain scattering data (one or more functions and/or special values of $\lambda$).
Then, just as in step 2 of the linear theory, one has explicit exponential evolution of the scattering data in time $t$.
In place of the inverse Fourier transform of the time-evolved transform data (step 3), one then has the \emph{inverse ST},  
which often requires the solution of a linear Riemann-Hilbert (RH) problem.

We illustrate these steps in a bit more detail for the defocusing NLS equation, namely \eqref{eq:NLS-general} with $\sigma=1$,
subject to initial data of the form \eqref{eq:schrod-initial-data}.
Eqn.~\eqref{eq:NLS-general} with $\sigma=1$ is the compatibility condition for the two linear equations of a \emph{Lax pair}:
\begin{equation}
\eps\frac{\partial{\bf w}}{\partial x}=\begin{bmatrix}-i\lambda & \psi\\\psi^{*} & i\lambda
\end{bmatrix}{\bf w},\quad\text{and}
\label{eq:defoc-nls-x-part}
\end{equation}
\begin{equation}
\eps\frac{\partial{\bf w}}{\partial t}=\begin{bmatrix}\displaystyle\vphantom{\frac{1}{1}}-i\lambda^2 -i\tfrac{1}{2}|\psi|^2 & \lambda\psi + i\tfrac{1}{2}\eps\psi_x\\
\displaystyle \vphantom{\frac{1}{1}}\lambda\psi^{*}-i\tfrac{1}{2}\eps\psi_x^{*} & i\lambda^2+i\tfrac{1}{2}|\psi|^2
\end{bmatrix}{\bf w},
\label{eq:defoc-nls-t-part}
\end{equation}
in which $\lambda\in\mathbb{C}$ is the spectral parameter.

Rigorous analysis of the initial-value problem for the defocusing NLS equation in the semiclassical limit may be carried out with the help of a suitable ST based on the scattering problem \eqref{eq:defoc-nls-x-part}.  The details are different depending on the type of boundary conditions that are enforced at $x=\pm\infty$.  

\subsubsection{The ST for rapidly decreasing boundary conditions}
Suppose first that the initial squared amplitude $\rho_0(x)$ is smooth and rapidly decreasing as $x\to\pm\infty$, say $\rho_0\in\mathscr{S}(\mathbb{R})$.  We wish to find the solution of \eqref{eq:NLS-general} with $\sigma=1$ that also decays for large $|x|$ for each $t>0$.  The direct ST involves calculating the \emph{Jost solution} $\mathbf{w}$ of the 
Zakharov-Shabat (ZS) equation \eqref{eq:defoc-nls-x-part} in which $t=0$ is fixed and $\psi$ is replaced by the initial data:
\begin{equation}
\eps\frac{d\mathbf{w}}{dx}=\begin{bmatrix}-i\lambda & \sqrt{\rho_0(x)}e^{iS(x)/\eps}\\
\sqrt{\rho_0(x)}e^{-iS(x)/\eps}& i\lambda\end{bmatrix}\mathbf{w},
\label{eq:Zakharov-Shabat}
\end{equation}
that is, the solution for $\lambda\in\mathbb{R}$ that is determined (assuming sufficiently rapid decay of $\rho_0$ for large $|x|$) by the conditions
\begin{equation}
\mathbf{w}(x)=\begin{cases}
\begin{bmatrix}e^{-i\lambda x/\eps}\\0\end{bmatrix} + R_0^\eps(\lambda)\begin{bmatrix}0\\e^{i\lambda x/\eps}\end{bmatrix}+o(1),&\quad x\to +\infty\\
T_0^\eps(\lambda)\begin{bmatrix}e^{-i\lambda x/\eps}\\0\end{bmatrix}+o(1),&\quad x\to -\infty
\end{cases}
\end{equation}
for some coefficients $R_0^\eps(\lambda)$ (the \emph{reflection coefficient}) and $T_0^\eps(\lambda)$ (the \emph{transmission coefficient}), satisfying $|R_0^\eps(\lambda)|^2+|T_0^\eps(\lambda)|^2=1$.

For the inverse ST, we solve (for each fixed $x$ and $t$) the following RH problem:  seek $\mathbf{M}:\mathbb{C}\setminus\mathbb{R}\to \mathrm{SL}(2,\mathbb{C})$ such that:
\begin{itemize}
\item \textbf{Analyticity}:  $\mathbf{M}$ is analytic in each half-plane, and takes boundary values $\mathbf{M}_\pm:\mathbb{R}\to \mathrm{SL}(2,\mathbb{C})$ on $\mathbb{R}$ from $\mathbb{C}_\pm$.
\item \textbf{Jump Condition}:  The boundary values are related by $\mathbf{M}_+(\lambda)=\mathbf{M}_-(\lambda)\mathbf{V}(\lambda)$ for $\lambda\in\mathbb{R}$, where the \emph{jump matrix} is defined by
\begin{equation}
\mathbf{V}(\lambda):=\begin{bmatrix}1-|R_0^\eps(\lambda)|^2 &
-e^{-2i(\lambda x+\lambda^2t)/\eps}R_0^\eps(\lambda)^*\\e^{2i(\lambda x+\lambda^2t)/\eps}R_0^\eps(\lambda) & 1\end{bmatrix}.
\end{equation}
\item \textbf{Normalization}:  As $\lambda\to\infty$, $\mathbf{M}(\lambda)\to\mathbb{I}$.
\end{itemize}
The solution of the initial-value problem is then given by 
\begin{equation}
\psi(x,t)=2i\lim_{\lambda\to\infty}\lambda M_{12}(\lambda).
\end{equation}

\subsubsection{The ST for finite density boundary conditions}  
A simple exact solution of the defocusing NLS equation is the $x$-independent \emph{background wave} $\psi=\psi_0(t):= e^{-it/\epsilon}$.  Suppose now that $\rho_0(x)\to 1$ and $S(x)\to 0$ as $x\to\pm\infty$.  We may then seek the solution of the initial-value problem for the defocusing NLS equation subject to the \emph{finite density} boundary condition that $\psi(x,t)\to\psi_0(t)$ as $x\to \pm\infty$ for each $t>0$.  

The presence of the background wave opens up a gap $(-1,1)$ in the continuous spectrum of the ZS equation \eqref{eq:Zakharov-Shabat}, and a Jost solution with oscillatory asymptotics as $x\to\pm\infty$ only exists for $\lambda\in \mathbb{R}\setminus (-1,1)$.  Analogues of the reflection and transmission coefficient are therefore defined for such $\lambda$.  On the other hand, in the spectral gap there may exist discrete eigenvalues, values of $\lambda$ for which there is a nonzero solution $\mathbf{w}(x)$ that decays rapidly for large $|x|$.  Labeling these in increasing order as $\lambda_1<\lambda_2<\cdots<\lambda_N$, each such eigenvalue is associated with a two-component eigenfunction $\mathbf{w}=\mathbf{w}_j(x)$ that is determined up to a sign\footnote{The sign is determined by the assertion that $\chi_j$ is real in \eqref{eq:chi-norm}.} by the properties
\begin{equation}
\mathbf{w}_j(x)^* = i\sigma_1\mathbf{w}_j(x)\quad\text{and}\quad \int_\mathbb{R}|w_j(x)|^2\,dx=1
\end{equation}
where in the latter equation the scalar $w_j(x)$ stands for either of the two components of $\mathbf{w}_j(x)$.  This allows a norming constant $\chi_j\in\mathbb{R}$ to be defined for each eigenvalue $\lambda_j$ by the formula
\begin{equation}
\mathbf{w}_j(x)=\left(\begin{bmatrix}ie^{-i\omega_j/2}\\-e^{i\omega_j/2}\end{bmatrix} + o(1)
\right)e^{(\chi_j-x\sin(\omega_j))/\epsilon},\quad x\to +\infty,
\label{eq:chi-norm}
\end{equation}
where  $\lambda_j=\cos(\omega_j)$ for $0<\omega_j<\pi$.  The calculation of the reflection coefficient and the discrete data $\{(\omega_j,\chi_j)\}_{j=1}^N$ constitute the direct ST in this case.

The inverse ST may again be phrased in terms of a RH problem, in general a more complicated one than in the case of rapidly-decreasing boundary conditions as it has to be formulated on a Riemann surface (albeit one of genus zero that  can be mapped to the complex plane by stereographic projection) and in general the matrix unknown has poles at the eigenvalues $\{\lambda_1,\dots,\lambda_N\}$ satisfying certain residue conditions that are part of the formulation of the problem.  However, in the special case that the reflection coefficient vanishes identically on the continuous spectrum $\mathbb{R}\setminus (-1,1)$, the matrix unknown $\mathbf{M}(\lambda)$ may be considered to be a meromorphic function of $\lambda$ with simple poles at the eigenvalues, and the residue conditions allow the RH problem to be solved via a partial-fractions ansatz, reducing the problem to one of linear algebra in dimension $N$.  Solving this problem by Cramer's rule, Jin, Levermore, and McLaughlin \cite{JinLM99} (following and simplifying the original approach of Zakharov and Shabat \cite{ZakharovS73}) showed in particular that 
\begin{equation}
\rho(x,t):=|\psi(x,t)|^2 = 1-\eps^2\frac{\partial^2}{\partial x^2}\log(\tau(x,t)),
\label{eq:defoc-nls-rho-from-tau}
\end{equation}
where the ``$\tau$-function'' is defined by
\begin{equation}
\tau(x,t):=\det(\mathbb{I}+\mathbf{G}(x,t))
\label{eq:defoc-nls-tau-function}
\end{equation}
with $\mathbf{G}(x,t)$ being the $N\times N$ matrix with elements
\begin{equation}
G_{jk}(x,t):=\frac{\eps e^{(\chi_j+\chi_k-(x+\cos(\omega_j)t)\sin(\omega_j)-(x+\cos(\omega_k)t)\sin(\omega_k))/\eps}}{2\sin(\tfrac{1}{2}(\omega_j+\omega_k))}.
\end{equation}

\subsubsection{Weak semiclassical limits via Lax-Levermore theory}
We now describe a general approach to singular asymptotics for integrable equations that was first developed by Lax and Levermore \cite{LaxL83} in the context of the zero-dispersion limit for the KdV equation.  Here we summarize aspects of this approach as it applies to the semiclassical limit for the defocusing NLS equation with finite-density boundary conditions \cite{JinLM99}.  As will be seen, the Lax-Levermore (LL) method provides an accurate description of the asymptotic behavior of the initial-value problem for \eqref{eq:NLS-general} with $\sigma=1$ in a weak topology that averages over any rapid oscillations that may appear in the solution.

Given that the small parameter $\eps$ appears in the initial data as well as explicitly in the direct spectral problem \eqref{eq:defoc-nls-x-part}, the first step is the analysis of the scattering data to determine its dependence on $\eps$ in the limit $\eps\to 0$.  It is convenient to assume that the initial data functions $\rho_0$ and $u_0$ are such that the combinations
\begin{equation}
\alpha(x):= -\frac{1}{2}u_0(x)-\sqrt{\rho_0(x)},\quad \beta(x):=-\frac{1}{2}u_0(x)+\sqrt{\rho_0(x)}
\label{eq:alpha-beta}
\end{equation}
are functions each with a single critical point, a maximizer $x_\alpha$ for $\alpha$ with value $\lambda^-:=\alpha(x_\alpha)$ and a minimizer $x_\beta$ for $\beta$ with value $\lambda^+:=\beta(x_\beta)$.  We also assume that $\lambda^-<\lambda^+$.  Note that as $\rho_0(x)\to 1$ and $u_0(x)\to 0$ as $x\to\pm\infty$, we have $\alpha(x)\to -1$ and $\beta(x)\to 1$ in this limit.  Under these conditions, the direct spectral problem admits a formal analysis for $\lambda\in\mathbb{R}$ in the limit $\eps\to 0$ by the WKB method.  Solutions are rapidly oscillatory (resp., exponential) whenever $(\lambda-\alpha(x))(\lambda-\beta(x))$ is positive (resp., negative).  In particular, whenever $|\lambda|>1$ solutions are oscillatory for all $x\in\mathbb{R}$, a fact that can be turned into a proof that the reflection coefficient defined for $|\lambda|\ge 1$ tends to zero with $\eps$ (the analysis is substantially more challenging if we consider $|\lambda|>1$ but $|\lambda|\approx 1$).  It remains to characterize the discrete spectrum in the interval $(-1,1)$.  If $\lambda^-<\lambda<\lambda^+$, then solutions are either exponentially growing or exponentially decaying for all $x\in\mathbb{R}$, a fact which suggests that no such $\lambda$ can be an eigenvalue.  However, if either $-1<\lambda<\lambda^-$ or $\lambda^+<\lambda<1$, then there exist precisely two turning points $x^-(\lambda)<x^+(\lambda)$ such that solutions are exponentially growing or decaying for $x<x^-(\lambda)$ and $x>x^+(\lambda)$ but are rapidly oscillatory for $x^-(\lambda)<x<x^+(\lambda)$.  In this situation one can derive connection formulae at the turning points that allow the WKB solutions to be continued through them, suggesting that 
with the phase integral defined by
\begin{equation}
\Phi(\lambda):=\int_{x^-(\lambda)}^{x^+(\lambda)}\sqrt{(\lambda-\alpha(x))(\lambda-\beta(x))}\,dx,
\label{eq:LL-phase-integral}
\end{equation}
the eigenvalues in the interval $-1<\lambda<\lambda^-$ (resp., in the interval $\lambda^+<\lambda<1$) are well-approximated by numbers $\tilde{\lambda}^{(\alpha)}_j$ (resp., $\tilde{\lambda}^{(\beta)}_j$) defined by the Bohr-Sommerfeld quantization rule $\Phi(\tilde{\lambda}^{(\alpha)}_j)=(j-\tfrac{1}{2})\pi\eps$ for $j=1,\dots,N^-(\eps)$ (resp., $\Phi(\tilde{\lambda}^{(\beta)}_j)=(j-\tfrac{1}{2})\pi\eps$ for $j=1,\dots,N^+(\eps)$), where
\begin{equation}
N^\pm(\eps):=\left\lceil\frac{1}{\pi\eps}\int_\mathbb{R}\sqrt{(1\mp\alpha(x))(1\mp\beta(x))}\,dx\right\rceil.
\end{equation}
The norming constant corresponding to an (approximate) eigenvalue $\tilde{\lambda}$ is itself approximated by
\begin{multline}
\tilde{\chi}(\tilde{\lambda}):=x^+(\tilde{\lambda})\sqrt{1-\tilde{\lambda}^2}\\{}+\int_{x^+(\tilde{\lambda})}^{+\infty}\left(\sqrt{1-\tilde{\lambda}^2}-\sqrt{(\tilde{\lambda}-\alpha(x))(\beta(x)-\tilde{\lambda})}\right)\,dx.
\label{eq:LL-norming-constant}
\end{multline}

Based on these formal considerations, one may neglect the reflection coefficient entirely and define a determinantal $\tau$-function $\tilde{\tau}(x,t)$ (cf., \eqref{eq:defoc-nls-tau-function}) based on the $N(\eps):=N^-(\eps)+N^+(\eps)$ approximate eigenvalues and corresponding approximate norming constants.  This function is certainly associated with \emph{some} (exact multi-soliton) solution $\tilde{\psi}(x,t)$ of the defocusing NLS equation, but it is not necessarily the solution of the initial-value problem because the scattering data has been replaced with an approximation.  At the heart of the LL method is a rigorous asymptotic analysis of the quantity $\eps^2\log(\tilde{\tau}(x,t))$ in the limit $\eps\to 0$ in which the matrix size ($N(\eps)\times N(\eps)$) grows without bound.  The key observation is that the Fredholm expansion
\begin{equation}
\tilde{\tau}(x,t):=\det(\mathbb{I}+\tilde{\mathbf{G}}(x,t))=\sum_{S\subset\{1,\dots,N(\eps)\}}\det(\tilde{\mathbf{G}}_S(x,t)),
\end{equation}
where $\tilde{\mathbf{G}}_S(x,t)$ is the square minor of $\tilde{\mathbf{G}}(x,t)$ with row/column indices taken from $S$ (by convention $\det(\tilde{\mathbf{G}}_\emptyset(x,t)):=1$), consists only of positive terms that can be expressed in closed-form.  The essence of the LL approach is to show that for $\eps>0$ sufficiently small the Fredholm expansion is well-approximated by its largest term, leading to a finite maximization problem.  We may think of a subset $S$ that maximizes $\det(\tilde{\mathbf{G}}_S(x,t))$ as a measure made of equal Dirac masses supported at the $\omega$-values in $(0,\pi)$ corresponding to selected approximate eigenvalues in $(-1,1)$, and therefore it makes sense that in the limit $\eps\to 0$ in which the eigenvalues fill out the intervals $(-1,\lambda^-)\cup (\lambda^+,1)$ with a continuous density $|\Phi'(\lambda)|$, the discrete maximization problem can in turn be approximated by a maximization problem involving a certain functional of a measure supported on corresponding subintervals of $[0,\pi]$.  In fact, it is enough to consider absolutely continuous measures with densities $\eta\in L^1(0,\pi)$.  To set up the limiting extremal problem, begin with the asymptotic (Weyl) density of angles $\omega\in (0,\pi)$ of eigenvalues, $\overline{\eta}(\omega):=(1-\chi_{(\lambda^-,\lambda^+)}(\cos(\omega)))|\Phi'(\cos(\omega))|\sin(\omega)$, and consider the admissible set $\mathcal{A}:=\{\eta\in L^1(0,\pi): 0\le \eta(\omega)\le \overline{\eta}(\omega)\}$.  For $\eta\in\mathcal{A}$ we define the functional
\begin{multline}
Q(\eta;x,t):=\frac{2}{\pi}\int_0^\pi\left[\tilde{\chi}(\cos(\omega))-(x+\cos(\omega)t)\sin(\omega)\right]\,\eta(\omega)\,d\omega \\{}+
\frac{1}{\pi^2}\int_0^\pi\int_0^\pi\ln\left(\left|\frac{\sin(\tfrac{1}{2}(\omega-\omega'))}{\sin(\tfrac{1}{2}(\omega+\omega'))}\right|\right)\eta(\omega)\eta(\omega')\,d\omega\,d\omega'.
\label{eq:JinLM-Q}
\end{multline}
The LL maximization problem for the semiclassical defocusing NLS equation is then the following:  given $(x,t)\in\mathbb{R}^2$, find the maximum value $q(x,t)$ of $Q(\eta;x,t)$ as $\eta$ ranges over the admissible set $\mathcal{A}$.  The main result \cite[Theorem 3.6]{JinLM99} is that
$\eps^2\log(\tilde{\tau}(x,t))$ converges to $q(x,t)$ as $\eps\to 0$, with the convergence being uniform on compact subsets of $\mathbb{R}^2$ and with the maximum $q(x,t)$ being a continuous function.
It can be proved that the functional $Q(\eta;x,t)$ is, for each $(x,t)\in\mathbb{R}^2$, strictly convex, i.e., for all $y\in (0,1)$,
\begin{equation}
Q(y\eta_1+(1-y)\eta_0;x,t)>yQ(\eta_1;x,t)+(1-y)Q(\eta_0;x,t),
\end{equation}
and that $Q$ is bounded above.  As the constraints on $\eta$ are linear, this is a well-posed extremal problem with a unique maximizer.

To extract information about the solution $\tilde{\psi}(x,t)$ of the defocusing NLS equation, one has to take derivatives of $\eps^2\log(\tilde{\tau}(x,t))$ (cf., \eqref{eq:defoc-nls-rho-from-tau}).  Now, the locally uniform convergence of $\eps^2\log(\tilde{\tau}(x,t))$ to $q(x,t)$ implies convergence in the sense of distributions, which commutes with differentiation.  Hence one concludes that $\tilde{\rho}(x,t):=|\tilde{\psi}(x,t)|^2$ converges in the sense of distributions to $1-q_{xx}(x,t)$ as $\eps\to 0$.  Because distributional convergence entails convergence of integrals against $\eps$-independent test functions, the limit process essentially replaces any oscillations on $o(1)$ wavelengths with their local averages; hence some fine-structure information is lost in the computation of the (weak) limit.  Nonetheless, the limit can be strengthened under certain conditions on $(x,t)$, in particular if $t$ is sufficiently small, and as such it can be shown that $\tilde{\rho}(x,0)\to \rho_0(x)$ and $\tilde{u}(x,0)\to u_0(x)$ in the (strong) $L^1(\mathbb{R})$ sense.  This latter result is taken as justification after the fact of the replacement of the true solution $\psi(x,t)$ of the initial-value problem  with the function $\tilde{\psi}(x,t)$ whose $\tau$-function is the object of study in LL theory.

\subsubsection{Strong semiclassical asymptotics by RH analysis}
Next we illustrate a different and more modern approach to singular asymptotics by considering the initial-value problem for \eqref{eq:NLS-general} with $\sigma=1$ in the semiclassical limit $\eps\to 0$ but subject to rapidly-decreasing boundary conditions.  To our knowledge, the details of the calculations to follow have not before appeared in the literature, although the ideas are not new.  The main tool is the Deift-Zhou (DZ) steepest descent method for RH problems first introduced in \cite{DeiftZ92,DeiftZ93} and subsequently extended and further developed by many authors.

\begin{figure}[h]
\begin{center}
\includegraphics[width=\linewidth]{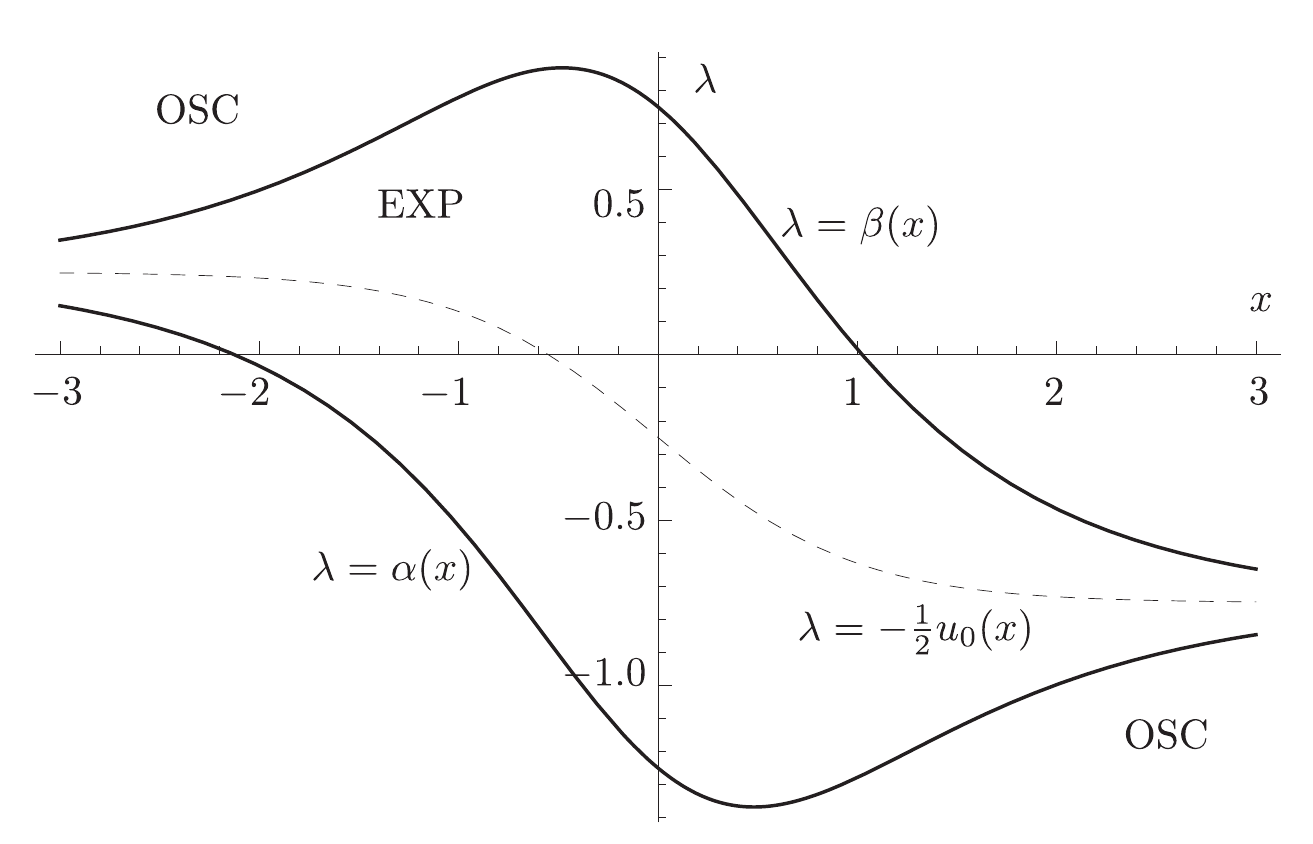}
\end{center}
\caption{The curves $\lambda=\alpha(x)$ and $\lambda=\beta(x)$ for rapidly decreasing initial data.}
\label{fig:TurningPoints}
\end{figure}

As in the LL method, the first step is to analyze the scattering data for the spectral problem \eqref{eq:defoc-nls-x-part} in the limit $\eps\to 0$.  Recalling the functions $\alpha(x)$ and $\beta(x)$ defined in terms of the initial data by \eqref{eq:alpha-beta}, we assume that the functions $u_0'(\cdot)$ and $\rho_0(\cdot)$ are, say, Schwartz-class functions for which $\alpha$ and $\beta$ are again functions each with only one critical point, a minimizer $x_\alpha$ for $\alpha$ with minimum value $\lambda^-:=\alpha(x_\alpha)$ and a maximizer $x_\beta$ for $\beta$ with maximum value $\lambda^+:=\beta(x_\beta)$ (see Fig.~\ref{fig:TurningPoints}).
In the WKB approximation, solutions of the ZS equation are rapidly oscillatory or exponentially growing or decaying in certain $\lambda$-dependent intervals of the $x$-axis as indicated in Fig.~\ref{fig:TurningPoints}.  These intervals are separated by exactly two turning points $x^-(\lambda)<x^+(\lambda)$ defined for $\lambda\in (\lambda^-,\lambda^+)$, while if either $\lambda<\lambda^-$ or $\lambda>\lambda^+$ there are no turning points and solutions of the ZS problem are rapidly oscillatory for all $x\in\mathbb{R}$.

The WKB formalism augmented with connection analysis at turning points yields the following results for the reflection coefficient $R_0^\eps(\lambda)$.  If $\lambda<\lambda^-$ or $\lambda>\lambda^+$ then one has an analogue of an ``above-barrier'' reflection problem, and $R_0^\eps(\lambda)=\bo(\eps)$.  On the other hand, if $\lambda\in (\lambda^-,\lambda^+)$, then 
\begin{equation}
R_0^\eps(\lambda)=e^{2i\theta_0(\lambda)/\eps}(1+\bo(\eps)),\quad\text{and}
\end{equation}
\begin{equation}
|T_0^\eps(\lambda)|^2=1-|R_0^\eps(\lambda)|^2=e^{-2m(\lambda)/\eps}(1+\bo(\eps)),
\end{equation}
where \begin{equation}
m(\lambda):=\int_{x^-(\lambda)}^{x^+(\lambda)}
\sqrt{(\lambda-\alpha(x))(\beta(x)-\lambda)}
\,dy
\label{eq:m-define}
\end{equation}
and with $\sigma:=\mathrm{sgn}(\lambda+\tfrac{1}{2}u_0(+\infty))$,
\begin{multline}
\theta_0(\lambda):=-\tfrac{1}{2}S(x^+(\lambda))-\lambda x^+(\lambda)\\
{}+\int_{x^+(\lambda)}^{+\infty}
\left[\sigma
\sqrt{(\lambda-\alpha(x))(\lambda-\beta(x))}
-(\lambda+\tfrac{1}{2}u_0(y))\right]\,dy.
\label{eq:theta0-define}
\end{multline}
Formulae \eqref{eq:m-define} and \eqref{eq:theta0-define} could be qualitatively compared with \eqref{eq:LL-phase-integral} and \eqref{eq:LL-norming-constant} respectively.  These calculations can be made completely rigorous with the use of Langer transformations to map the differential equation \eqref{eq:defoc-nls-x-part} onto a sufficiently controllable perturbation of the Airy equation on intervals containing exactly one turning point.  Such analysis fails for $\lambda\approx \lambda^\pm$ where the turning points $x^\pm(\lambda)$ collide.  Nonetheless, we will proceed as in the LL approach and simply replace the actual reflection coefficient $R_0^\eps(\lambda)$ with an approximation defined by
\begin{equation}
\tilde{R}_0^\eps(\lambda):=\chi_{[\lambda^-,\lambda^+]}(\lambda)H^\eps(\lambda)e^{2i\theta_0(\lambda)/\eps},\quad \lambda\in\mathbb{R},
\end{equation}
where $H^\eps(\lambda):=\sqrt{1-e^{-2m(\lambda)/\eps}}$.


The idea is to now take $\tilde{R}_0^\eps(\cdot)$ as the reflection coefficient associated with some (unknown) initial data, formulate the RH problem of inverse scattering for the corresponding matrix $\tilde{\mathbf{M}}(\lambda)$, and to rigorously analyze this problem in the semiclassical limit $\eps\to 0$.  As in the LL method, one expects to be able to show that at $t=0$ the extracted solution $\tilde{\psi}(x,t)$ (which by means of the dressing method can be shown to be an exact solution of the defocusing NLS equation) is suitably close to the true initial data $\psi(x,0)$. This obviously requires being able to solve matrix RH problems involving small parameters, so we pause to summarize the necessary background.

\paragraph{Aside:  solution of RH problems}
Let $\Sigma$ be an oriented contour (perhaps with self-intersection points), and let $\mathbf{V}:\Sigma\to\mathrm{SL}(2,\mathbb{C})$ be a given jump matrix decaying to $\mathbb{I}$ as $\lambda\to\infty$ along any unbounded arcs of $\Sigma$.  A general RH problem is the following:  find $\mathbf{M}:\mathbb{C}\setminus\Sigma\to\mathrm{SL}(2,\mathbb{C})$ such that:
\begin{itemize}
\item\textbf{Analyticity}:  $\mathbf{M}$ is analytic in its domain of definition, and takes boundary values $\mathbf{M}_\pm:\Sigma\to\mathrm{SL}(2,\mathbb{C})$ on $\Sigma$ from the left ($+$) and
right ($-$).
\item\textbf{Jump Condition}:  The boundary values are related by $\mathbf{M}_+(\lambda)=\mathbf{M}_-(\lambda)\mathbf{V}(\lambda)$ for $\lambda\in\Sigma$ (avoiding any self-intersection points).
\item\textbf{Normalization}:  As $\lambda\to\infty$, $\mathbf{M}(\lambda)\to\mathbb{I}$.
\end{itemize}
This problem can be studied by converting it into a linear system of singular integral equations as follows.
Subtracting $\mathbf{M}_-(\lambda)$ from both sides of the jump condition yields
\begin{equation}
\mathbf{M}_+(\lambda)-\mathbf{M}_-(\lambda)=\mathbf{M}_-(\lambda)(\mathbf{V}(\lambda)-\mathbb{I}),\quad \lambda\in\Sigma.
\end{equation}
Therefore, taking into account the analyticity of $\mathbf{M}$ in $\mathbb{C}\setminus\Sigma$ and the asymptotic value of $\mathbb{I}$ as $\lambda\to\infty$ it is necessary that $\mathbf{M}(\lambda)$ is given by the Cauchy integral (Plemelj formula):
\begin{equation}
\mathbf{M}(\lambda)=\mathbb{I}+\frac{1}{2\pi i}\int_\Sigma\frac{\mathbf{M}_-(\mu)(\mathbf{V}(\mu)-\mathbb{I})}{\mu-\lambda}\,d\mu,\quad \lambda\in\mathbb{C}\setminus\Sigma.
\end{equation}
Letting $\lambda$ tend to $\Sigma$ from the right (we denote this by $\lambda_-$) we obtain a closed equation for the boundary value $\mathbf{M}_-(\lambda)$:
\begin{equation}
\mathbf{X}(\lambda)-\frac{1}{2\pi i}\int_\Sigma\frac{\mathbf{X}(\mu)(\mathbf{V}(\mu)-\mathbb{I})}{\mu-\lambda_-}\,d\mu = \frac{1}{2\pi i}\int_\Sigma\frac{\mathbf{V}(\mu)-\mathbb{I}}{\mu-\lambda_-}\,d\mu,
\label{eq:singular-integral-equation}
\end{equation}
where $\lambda\in\Sigma$ and $\mathbf{X}(\lambda):=\mathbf{M}_-(\lambda)-\mathbb{I}$.  

If the jump matrix $\mathbf{V}$ depends on parameters (e.g., $x$, $t$, $\eps$), one can consider the asymptotic behavior of the RH problem with respect to one or more parameters.
While one could attempt to analyze the singular integral equation \eqref{eq:singular-integral-equation}, this would generally be a difficult (perhaps impossible) task.  Indeed, in general it is wiser to keep in mind the complex-analytic origin of this equation and to work with the RH problem itself.  This is the main motivation behind the DZ steepest descent method.

On the other hand, the singular integral equation \eqref{eq:singular-integral-equation} is perhaps the most useful in the \emph{small norm setting}.  This means that $\mathbf{V}-\mathbb{I}\in L^2(\Sigma)\cap L^\infty(\Sigma)$ and is small in the $L^\infty(\Sigma)$ sense.  The utility of such estimates is a consequence of the fact 
that for a general class of contours $\Sigma$, the operator
\begin{equation}
\mathbf{F}\mapsto\frac{1}{2\pi i}\int_\Sigma\frac{\mathbf{F}(\mu)\,d\mu}{\mu-\lambda_-}
\end{equation}
is bounded on $L^2(\Sigma)$, with a norm that only depends on geometrical details of $\Sigma$.
This was first proven in the case that $\Sigma$ is a Lipschitz arc by McIntosh, Coifman, and Meyer \cite{McIntoshCM82} and a trick necessary to extend the result to contours $\Sigma$ with ``reasonable'' self-intersections is explained, for example, in \cite[Lemma 8.1]{BealsC84}.
For problems of small-norm type, the singular integral equation \eqref{eq:singular-integral-equation} can be solved in $L^2(\Sigma)$ by iteration (for an elementary description, see \cite[Appendix B]{BuckinghamM14}).  From the abstract point of view, the convergence of the iterates 
guarantees existence and uniqueness of the solution.  However from the practical point of view it
also allows the solution to be constructed (approximated with arbitrary accuracy and estimated).  The $L^2(\Sigma)$ norm of $\mathbf{X}$ is proportional to that of $\mathbf{V}-\mathbb{I}$.  The last point is that 
under suitable other technical assumptions, $\mathbf{M}(\lambda)$ has an asymptotic expansion as $\lambda\to\infty$:
\begin{equation}
\mathbf{M}(\lambda)=\mathbb{I}+\sum_{n=1}^N\lambda^{-n}\mathbf{M}_n + \bo(\lambda^{-(N+1)}),\quad\lambda\to\infty
\end{equation}
and the moments $\mathbf{M}_n$ are bounded in terms of norms of $\mathbf{V}-\mathbb{I}$.

\paragraph{The $g$-function mechanism}
Now we return to the specific RH problem at hand:
Seek $\tilde{\mathbf{M}}:\mathbb{C}\setminus [\lambda^-,\lambda^+]\to \mathrm{SL}(2,\mathbb{C})$ with the following properties:
\begin{itemize}
\item\textbf{Analyticity}:  $\tilde{\mathbf{M}}$ is analytic in its domain of definition and takes boundary values $\tilde{\mathbf{M}}_\pm(\lambda)$ on $(\lambda^-,\lambda^+)$ from $\mathbb{C}_\pm$.
\item\textbf{Jump Condition}:  $\tilde{\mathbf{M}}_+(\lambda)=\tilde{\mathbf{M}}_-(\lambda)\tilde{\mathbf{V}}(\lambda)$ for $\lambda^-<\lambda<\lambda^+$, where
\begin{equation}
\tilde{\mathbf{V}}(\lambda):=\begin{bmatrix}e^{-2m(\lambda)/\eps} & 
-e^{-2i\theta(\lambda;x,t)/\eps}H^\eps(\lambda)\\
e^{2i\theta(\lambda;x,t)/\eps}H^\eps(\lambda)& 1\end{bmatrix},
\end{equation}
with $\theta(\lambda;x,t):=\lambda x+\lambda^2t+\theta_0(\lambda)$.
\item\textbf{Normalization}:  As $\lambda\to\infty$, $\tilde{\mathbf{M}}(\lambda)\to\mathbb{I}$.
\end{itemize}

The key idea introduced by Deift, Venakides, and Zhou (first in \cite{DeiftVZ94}, and then generalized to a setting similar to the present one in \cite{DeiftVZ97}) is that of a ``$g$-function''.  
Let $g:\mathbb{C}\setminus [\lambda^-,\lambda^+]\to\mathbb{C}$ be analytic with $g(\infty)=0$, and make the substitution $\tilde{\mathbf{M}}(\lambda)=\mathbf{N}(\lambda)e^{ig(\lambda)\sigma_3/\eps}$. Then $\mathbf{N}:\mathbb{C}\setminus [\lambda^-,\lambda^+]\to \mathrm{SL}(2,\mathbb{C})$ satisfies the conditions of this related RH problem:
\begin{itemize}
\item\textbf{Analyticity}:  $\mathbf{N}$ is analytic in $\mathbb{C}\setminus[\lambda^-,\lambda^+]$, taking boundary values $\mathbf{N}_\pm(\lambda)$ on $[\lambda^-,\lambda^+]$ from $\mathbb{C}_\pm$.
\item\textbf{Jump Condition}:  The boundary values are related by $\mathbf{N}_+(\lambda)=\mathbf{N}_-(\lambda)\mathbf{V}^{(\mathbf{N})}(\lambda)$ for $\lambda^-<\lambda<\lambda^+$, where 
\begin{equation}
\mathbf{V}^{(\mathbf{N})}(\lambda):=
\begin{bmatrix}
e^{2(\Delta(\lambda)-m(\lambda))/\eps} & -e^{-2i\phi(\lambda)/\eps}H^\eps(\lambda)\\
e^{2i\phi(\lambda)/\eps}H^\eps(\lambda) & e^{-2\Delta(\lambda)/\eps}\end{bmatrix},
\end{equation}
with 
$2\Delta(\lambda):=-i(g_+(\lambda)-g_-(\lambda))$ and $2\phi(\lambda):=2\theta(\lambda)-g_+(\lambda)-g_-(\lambda)$.
\item\textbf{Normalization}:  As $\lambda\to\infty$, $\mathbf{N}(\lambda)\to\mathbb{I}$.
\end{itemize}
We further suppose that $g(\lambda)=g(\lambda^*)^*$, making $\phi$ and $\Delta$ real.

The $g$-function is otherwise free to be chosen.  Suppose that by choice of $g$ it can be arranged that $(\lambda^-,\lambda^+)$ splits into three types of subintervals:
\begin{itemize}
\item \textbf{Voids}:  These are characterized by the conditions $\Delta(\lambda)\equiv 0$ and $\phi'(\lambda)>0$.
\item \textbf{Bands}:  These are characterized by the conditions $0<\Delta(\lambda)<m(\lambda)$
and $\phi'(\lambda)\equiv 0$.
\item\textbf{Saturated regions}:  These are characterized by the conditions $\Delta(\lambda)\equiv m(\lambda)$ and $\phi'(\lambda)<0$.
\end{itemize}
We sometimes collectively refer to voids and saturated regions as \emph{gaps}.  We always assume that gaps are separated by bands, and that the left and right-most subintervals of $(\lambda^-,\lambda^+)$ are gaps.
We now examine the consequences of each type of interval for the jump matrix $\mathbf{V}^{(\mathbf{N})}(\lambda)$.

\paragraph{Voids}
Under the condition that $\Delta(\lambda)\equiv 0$, the jump matrix $\mathbf{V}^{(\mathbf{N})}(\lambda)$ has an ``upper-lower'' factorization:
\begin{equation}
\mathbf{V}^{(\mathbf{N})}(\lambda)=\mathbf{U}(\lambda)\mathbf{L}(\lambda),
\label{eq:void-factorize}
\end{equation}
where
\begin{equation}
\mathbf{U}(\lambda):=\begin{bmatrix}1 & -e^{-2i\phi(\lambda)/\eps}
H^\eps(\lambda)\\0 & 1\end{bmatrix},\;
\mathbf{L}(\lambda):=
\begin{bmatrix}1 & 0\\e^{2i\phi(\lambda)/\eps}H^\eps(\lambda) & 1\end{bmatrix}.
\label{eq:UpperLowerFactors}
\end{equation}
Let us assume for simplicity that $\phi(\lambda)$ and $H^\eps(\lambda)$ are analytic functions\footnote{If  the initial data functions $u_0$ and $\rho_0$ are real analytic, then $m$ is analytic in $(\lambda^-,\lambda^+)$ except at the points $\lambda=-\tfrac{1}{2}u_0(\pm\infty)$, $\theta_0$ is analytic except at the point $\lambda=-\tfrac{1}{2}u_0(+\infty)$, and on intervals not containing the point $\lambda=-\tfrac{1}{2}u_0(-\infty)$, $\theta_0\pm im$ is locally the boundary value of a function analytic in $\mathbb{C}_\pm$.  It follows that the function $\phi$ is analytic in voids that do not contain the point $\lambda=-\tfrac{1}{2}u_0(+\infty)$ (because $\phi(\lambda)=\theta(\lambda)$ in voids) as well as in saturated regions that do not contain the point $\lambda=-\tfrac{1}{2}u_0(-\infty)$ (because $\phi(\lambda)=\theta(\lambda)\pm im(\lambda)-g_\pm(\lambda)$ in saturated regions).  It turns out that if $H^\eps$ is not analytic in a void or saturated region, the factorizations  \eqref{eq:void-factorize} and \eqref{eq:sat-factorize} can be replaced by three-factor factorizations in which the ``outer'' factors are as written in these formulae except that $H^\eps(\lambda)$ is replaced by $1$ and the ``inner'' factor compensates for the difference.  The inner factor is then exponentially close to $\mathbb{I}$ without any need to deform from the real line because $m(\lambda)>0$ making $H^\eps(\lambda)-1$ exponentially small.  On the other hand, analyticity of $H^\eps$ near $\lambda^\pm$ is more essential.} in the void.  Then the condition $\phi'(\lambda)>0$ makes $\phi(\lambda)$ a real 
analytic function that is strictly increasing in the void interval.  By the Cauchy-Riemann equations, 
it follows that the imaginary part of $\phi(\lambda)$ is positive (negative) in the upper (lower) half-plane.
Hence the first (second) matrix factor in \eqref{eq:void-factorize} has an analytic continuation into the lower (upper) half-plane that is exponentially close to the identity matrix in the limit $\eps\to 0$.

\paragraph{Bands}
The strict inequalities $0<\Delta(\lambda)<m(\lambda)$ imply that the diagonal elements of $\mathbf{V}^{(\mathbf{N})}(\lambda)$, namely
$e^{2(\Delta(\lambda)-m(\lambda))/\eps}$ and $e^{-2\Delta(\lambda)/\eps}$,
are both exponentially small in the semiclassical limit $\eps\to 0$.  The condition $\phi'(\lambda)\equiv 0$ together with the inequality $m(\lambda)>0$ that holds for all $\lambda\in (\lambda^-,\lambda^+)$ then implies that $\mathbf{V}^{(\mathbf{N})}(\lambda)$ is exponentially close in the semiclassical limit to a constant off-diagonal matrix:
\begin{equation}
\mathbf{V}^{(\mathbf{N})}(\lambda)=\begin{bmatrix}0 & -e^{-2i\phi/\eps}\\e^{2i\phi/\eps} & 0\end{bmatrix}+\text{exponentially small terms}.
\label{eq:band-approximate}
\end{equation}
The real constant $\phi$ can be different for different bands, and it generally can depend on $x$ and $t$ (but not $\eps$).  

\paragraph{Saturated regions}
Under the condition that $\Delta(\lambda)\equiv m(\lambda)$, the jump matrix $\mathbf{V}^{(\mathbf{N})}(\lambda)$ has a ``lower-upper'' factorization:
\begin{equation}
\mathbf{V}^{(\mathbf{N})}(\lambda)=\mathbf{L}(\lambda)\mathbf{U}(\lambda),
\label{eq:sat-factorize}
\end{equation}
with the factors being given by \eqref{eq:UpperLowerFactors}.
Again assuming for simplicity the analyticity of $\phi(\lambda)$ and $H^\eps(\lambda)$ in the saturated region, 
the condition $\phi'(\lambda)<0$ makes $\phi(\lambda)$ a real 
analytic function that is strictly decreasing.  By the Cauchy-Riemann equations, 
it follows that the imaginary part of $\phi(\lambda)$ is negative (positive) in the upper (lower) half-plane.
This again implies that the first (second) matrix factor in \eqref{eq:sat-factorize} has an analytic continuation into the lower (upper) half-plane that is exponentially close to the identity matrix in the limit $\eps\to 0$.

\paragraph{Aside:  variational meaning of $g$.  Connection with LL theory}
The three types of conditions on the boundary values of the function $g(\cdot)$ can be formulated as the Euler-Lagrange variational conditions for a certain functional maximization problem.  We follow, \emph{mutatis mutandis}, the approach of \cite[Section 6]{DeiftVZ97}.  The first step is to find a conformal mapping taking the domain of a priori analyticity of $g$, namely $\mathbb{C}\setminus [\lambda^-,\lambda^+]$, onto the open upper half $z$-plane.  We choose 
\begin{equation}
z(\lambda):=i\frac{(\lambda-\lambda^+)^{1/2}}{(\lambda-\lambda^-)^{1/2}}\;\text{with inverse}\;
\lambda(z):=\frac{\lambda^-z+\lambda^+z^{-1}}{z+z^{-1}}.
\end{equation}
Thus $\lambda=\lambda^+$ corresponds to $z=0$, $\lambda=\lambda^-$ corresponds to $z=\infty$, and $\lambda=\infty$ corresponds to $z=i$.  Also, given a value $\lambda\in (\lambda^-,\lambda^+)$, the boundary value $\lambda_-$ corresponds to a boundary point $z_+$ with $z>0$, while $\lambda_+$ then corresponds to $(-z)_+$.  Therefore, if we define a function $G(z)$ by setting
$G(z):=\mathrm{sgn}(\I\{z\})g(\lambda(z))$, then $G$ is an odd function of $z$ analytic for $z\in\mathbb{C}\setminus\mathbb{R}$, and that $G(\pm i)=0$ (from $g(\infty)=0$).  Since by assumption a right-neighborhood of $\lambda^-$ is a gap, and since $m(\lambda^-)=0$, the value $g(\lambda^-)$ is well-defined so $G(z)$ converges to opposite finite limits $G^\pm$ as $z\to\infty$ in $\mathbb{C}_\pm$.  Moreover, $G'(z)\to 0$ as $z\to\infty$.

The point of this transformation is that the boundary values $G_\pm(z)$ taken at a real $z$ from $\mathbb{C}_\pm$ are related to $g_\pm(\lambda(z))$ as follows:  $G_+(z)-G_-(z)=g_+(\lambda(z))+g_-(\lambda(z))$ and $G_+(z)+G_-(z)=-\mathrm{sgn}(z)(g_+(\lambda(z))-g_-(\lambda(z)))$.  That is, sums and differences of boundary values have been exchanged.  This is an important step in arriving at a variational problem with a logarithmic interaction as will be seen directly.  Another technique for exchanging certain sums and differences of boundary values will be discussed in \S\ref{sec:construction-of-g}.

Recalling the definition of the function $\phi(\cdot)$ in terms of the boundary values of $g$ we take a derivative and therefore, 
\begin{equation}
\begin{split}
G_+'(z)-G_-'(z)&=\frac{d}{dz}\left[g_+(\lambda(z))+g_-(\lambda(z))\right]\\
&=
2[\theta'(\lambda(z))-\phi'(\lambda(z))]\lambda'(z),\quad\forall z\in\mathbb{R}.
\end{split}
\end{equation}
Since $G'(z)\to 0$ as $z\to\infty$, $G'(z)$ is necessarily given in terms of the difference of its boundary values by the Plemelj formula:
\begin{equation}
G'(z)=\frac{1}{\pi i}\int_\mathbb{R}\frac{\theta'(\lambda(\zeta))-\phi'(\lambda(\zeta))}{\zeta-z}\lambda'(\zeta)\,d\zeta,\quad z\in\mathbb{C}\setminus\mathbb{R}.
\label{eq:Big-G-prime}
\end{equation}
As $[\theta'(\lambda(\cdot))-\phi'(\lambda(\cdot)]\lambda'(\cdot)$ is an odd function in $L^1(\mathbb{R})$, this formula confirms that $G'(z)$ is integrable at $z=\infty$.  From \eqref{eq:Big-G-prime}, $G(z)$ may then be obtained in each half-plane $\mathbb{C}_\pm$ separately as a contour integral with initial point $z=\pm i$, building in the condition $G(\pm i)=0$.  Therefore, for $z\in\mathbb{C}_\pm$ we have $G(z)=G_0(z)-G_0(\pm i)$, where
\begin{equation}
G_0(z)=-\frac{1}{\pi i}\int_\mathbb{R}[\theta'(\lambda(\zeta))-\phi'(\lambda(\zeta))]\lambda'(\zeta)\log(z-\zeta)\,d\zeta,
\end{equation}
in which $\log(\cdot)$ denotes the principal branch of the complex logarithm.  
Again using the fact that $[\theta'(\lambda(\cdot))-\phi'(\lambda(\cdot))]\lambda'(\cdot)$ is odd and assuming now that $\I\{z\}>0$ gives
\begin{equation}
G_0(z)=-\frac{1}{\pi i}\int_{\mathbb{R}_+}[\theta'(\lambda(\zeta))-\phi'(\lambda(\zeta))]\log\left(\frac{z-\zeta}{z+\zeta}\right)\lambda'(\zeta)\,d\zeta,
\end{equation}
again with the principal branch intended.
Mapping $\zeta>0$ back to the lower edge of the branch cut $[\lambda^-,\lambda^+]$ by $\zeta=\sqrt{\lambda^+-\mu}/\sqrt{\mu-\lambda^-}$ and writing $z=z(\lambda)$ we recover $g(\lambda)=G_0(z(\lambda))-G_0(i)$ in the form
\begin{equation}
g(\lambda)=\frac{1}{\pi i}\int_{\lambda^-}^{\lambda^+}[\theta'(\mu)-\phi'(\mu)]\left[K(\lambda,\mu)-K(\infty,\mu)\right]
\,d\mu.
\end{equation}
where for $\lambda\in\mathbb{C}\setminus [\lambda^-,\lambda^+]$ and $\mu\in (\lambda^-,\lambda^+)$,
\begin{equation}
K(\lambda,\mu):=
\log\left(\frac{i\sqrt{\mu-\lambda^-}(\lambda-\lambda^+)^{1/2}-\sqrt{\lambda^+-\mu}(\lambda-\lambda^-)^{1/2}}{i\sqrt{\mu-\lambda^-}(\lambda-\lambda^+)^{1/2}+\sqrt{\lambda^+-\mu}(\lambda-\lambda^-)^{1/2}}\right).
\end{equation}
Note that on this domain of definition, $0<\I\{K(\lambda,\mu)\}<\pi$.  Now, observe that upon letting $\lambda$ tend to $(\lambda^-,\lambda^+)$ from $\mathbb{C}_\pm$,
\begin{equation}
g_+(\lambda)-g_-(\lambda) = \frac{1}{\pi i}\int_{\lambda^-}^{\lambda^+}[\theta'(\mu)-\phi'(\mu)][K(\lambda_+,\mu)-K(\lambda_-,\mu)]\,d\mu,
\end{equation}
and the difference of boundary values of $K(\cdot,\mu)$ is given by
\begin{multline}
K(\lambda_+,\mu)-K(\lambda_-,\mu)=\\
-2\ln\left(\left|\frac{\sqrt{\mu-\lambda^-}\sqrt{\lambda^+-\lambda}-
\sqrt{\lambda^+-\mu}\sqrt{\lambda-\lambda^-}}
{\sqrt{\mu-\lambda^-}\sqrt{\lambda^+-\lambda}+
\sqrt{\lambda^+-\mu}\sqrt{\lambda-\lambda^-}}\right|\right)
\end{multline}
where now $\lambda$ and $\mu$ both lie in $(\lambda^-,\lambda^+)$.  Now mapping this interval to an angle $\omega\in (0,\pi)$ by $\lambda=\ell(\omega):=\tfrac{1}{2}((\lambda^++\lambda^-)+(\lambda^+-\lambda^-)\cos(\omega))$,
\begin{multline}
g_+(\ell(\omega))-g_-(\ell(\omega))=\\
\frac{2}{\pi i}\left[\int_0^\pi\eta(\omega')\ln\left(\left|\frac{\sin(\tfrac{1}{2}(\omega-\omega'))}{\sin(\tfrac{1}{2}(\omega+\omega'))}\right|\right)\,d\omega' + \pi a(\omega)\right],
\end{multline}
where $\eta(\omega):= \tfrac{1}{2}(\lambda^+-\lambda^-)\phi'(\ell(\omega))\sin(\omega)$ and where
\begin{multline}
a(\omega):=\\-\frac{\lambda^+-\lambda^-}{2\pi}\int_0^\pi\theta'(\ell(\omega'))\sin(\omega')
\ln\left(\left|\frac{\sin(\tfrac{1}{2}(\omega-\omega'))}{\sin(\tfrac{1}{2}(\omega+\omega'))}\right|\right)\,d\omega'.
\end{multline}
Since $g_+(\lambda)-g_-(\lambda)=2i\Delta(\lambda)$, we then see that the conditions that $\lambda=\ell(\omega)$ lies in a void, band, or saturated region amount to the following conditions on the (unknown) function $\eta(\omega)$ defined for $0<\omega<\pi$:
\begin{itemize}
\item\textbf{Voids:}  $\lambda=\ell(\omega)$ lies in a void if $\eta(\omega)>0$ and
\begin{equation}
\int_0^\pi\eta(\omega')\ln\left(\left|\frac{\sin(\tfrac{1}{2}(\omega-\omega'))}{\sin(\tfrac{1}{2}(\omega+\omega'))}\right|\right)\,d\omega' + \pi a(\omega)=0.
\end{equation}
\item\textbf{Bands:}  $\lambda=\ell(\omega)$ lies in a band if $\eta(\omega)=0$,
\begin{equation}
\int_0^\pi\eta(\omega')\ln\left(\left|\frac{\sin(\tfrac{1}{2}(\omega-\omega'))}{\sin(\tfrac{1}{2}(\omega+\omega'))}\right|\right)\,d\omega' + \pi a(\omega)<0,
\end{equation}
and
\begin{multline}
\int_0^\pi\eta(\omega')\ln\left(\left|\frac{\sin(\tfrac{1}{2}(\omega-\omega'))}{\sin(\tfrac{1}{2}(\omega+\omega'))}\right|\right)\,d\omega' + \pi a(\omega)\\
{}+2\pi m(\ell(\omega))>0.
\end{multline}
\item\textbf{Saturated regions:}  $\lambda=\ell(\omega)$ lies in a saturated region if $\eta(\omega)<0$ and
\begin{multline}
\int_0^\pi\eta(\omega')\ln\left(\left|\frac{\sin(\tfrac{1}{2}(\omega-\omega'))}{\sin(\tfrac{1}{2}(\omega+\omega'))}\right|\right)\,d\omega' + \pi a(\omega)\\{}+2\pi m(\ell(\omega))=0.
\end{multline}
\end{itemize}
It is then an exercise in the calculus of variations to check that these conditions on the unknown real-valued function $\eta\in L^1(0,\pi)$ are precisely the Euler-Lagrange variational conditions for the problem of maximizing over the whole space $L^1(0,\pi)$ the functional
\begin{multline}
Q(\eta;x,t):=\frac{2}{\pi}\int_0^\pi a(\omega)\eta(\omega)\,d\omega +\frac{2}{\pi}\int_0^\pi 2m(\ell(\omega))[\eta(\omega)]_-\,d\omega\\
{}+\frac{1}{\pi^2}\int_0^\pi\int_0^\pi\ln\left(\left|\frac{\sin(\tfrac{1}{2}(\omega-\omega'))}{\sin(\tfrac{1}{2}(\omega+\omega'))}\right|\right)\eta(\omega)\eta(\omega')\,d\omega\,d\omega',
\label{eq:g-function-Q}
\end{multline}
where $[\eta(\omega)]_-$ denotes the negative part of the function $\eta(\omega)$, i.e., $[\eta(\omega)]_-=\eta(\omega)$ if $\eta(\omega)<0$ and $[\eta(\omega)]_-=0$ otherwise.  The dependence of $Q$ on $(x,t)\in\mathbb{R}^2$ is linear and enters through the function $a(\omega)$, which in turn depends linearly on $\theta'(\cdot)$.  Using the fact that $m(\ell(\omega))>0$ it can be checked that the term in $Q$ involving $[\eta(\omega)]_-$ is convex, so by similar arguments as in LL theory the maximization problem is well-posed and has a unique solution.
Whether the maximizer $\eta$ has the property that it is zero or of a definite sign on interleaving intervals (bands or gaps respectively) is
a deeper regularity question that we will avoid later by a direct construction of the function $g$.
Nonetheless, the fact that the function $g$ has a variational characterization shows that the LL theory is in the background when one employs the DZ steepest descent technique.  Moreover, we will see soon that from this point of view the DZ method can be viewed as a technique for converting the weak asymptotics derived directly from the variational problem into strong asymptotics for the function $\tilde{\psi}(x,t)$.

\begin{figure}[h]
\begin{center}
\includegraphics[width=\linewidth]{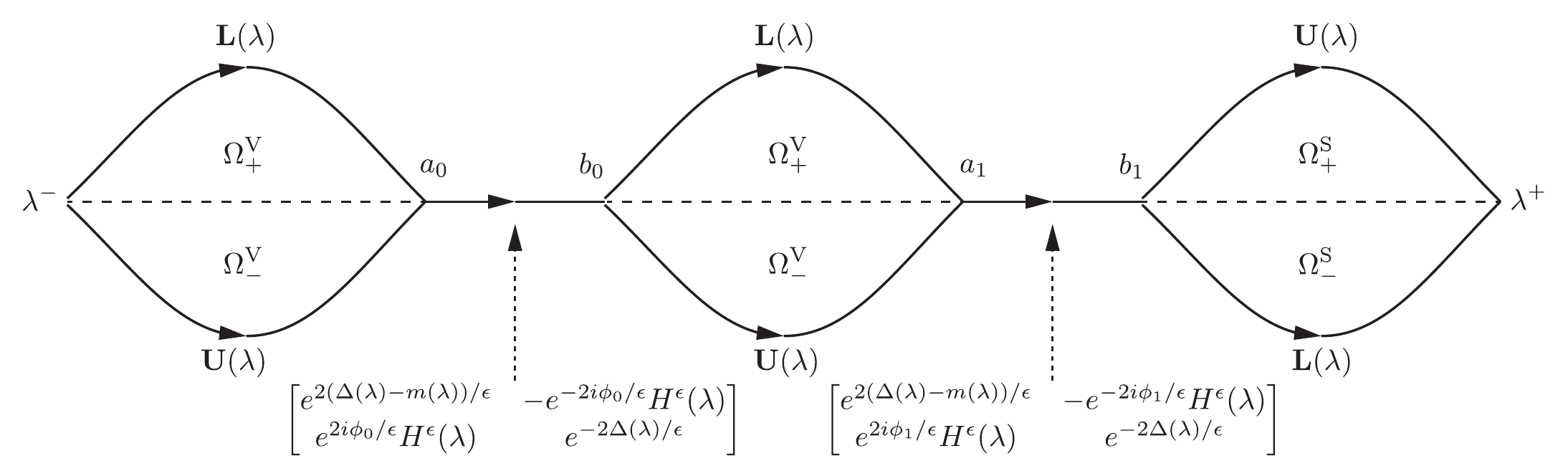}
\end{center}
\caption{The contour $\Sigma^{(\mathbf{O})}$ and jump matrix $\mathbf{V}^{(\mathbf{O})}$ for the ``opened lenses'' RH problem.}
\label{fig:RHP-O}
\end{figure}

\paragraph{Steepest descent}
To exploit the matrix factorizations, let $\Omega^\mathrm{V}_\pm$ ($\Omega^\mathrm{S}_\pm$) denote the union of thin lens-shaped domains in $\mathbb{C}_\pm$ that abut voids (saturated regions) as illustrated in Fig.~\ref{fig:RHP-O}.  
Recalling \eqref{eq:UpperLowerFactors}, define the piecewise analytic\footnote{Here we are again assuming for simplicity the analyticity of $\phi$ and $H^\eps$ in each gap.  More generally, even for analytic initial data, to maintain analyticity of $\mathbf{T}$ it may be necessary to omit the factor $H^\eps(\lambda)\approx 1$ from the off-diagonal matrix elements in lenses abutting certain gaps (or perhaps only in sub-domains of the indicated lenses --- we always retain the factor $H^\eps(\lambda)$ near the endpoints $\lambda^\pm$ for technical reasons).} matrix function $\mathbf{T}$ by
\begin{equation}
\mathbf{T}(\lambda):=\begin{cases}
\mathbf{L}(\lambda),
&\quad \lambda\in\Omega^\mathrm{V}_+,\\
\mathbf{L}(\lambda)^{-1},
&\quad\lambda\in\Omega^\mathrm{S}_-,\\
\mathbf{U}(\lambda)^{-1},
&\quad\lambda\in\Omega^\mathrm{V}_-\\
\mathbf{U}(\lambda),
&\quad\lambda\in\Omega^\mathrm{S}_+\\
\mathbb{I},&\quad\text{otherwise}.
\end{cases}
\end{equation}
Making the substitution $\mathbf{N}(\lambda)=\mathbf{O}(\lambda)\mathbf{T}(\lambda)$, one checks that $\mathbf{O}(\lambda)$ satisfies the following ``opened lenses'' RH problem:
\begin{itemize}
\item\textbf{Analyticity}: $\mathbf{O}$ is analytic in $\mathbb{C}\setminus\Sigma^{(\mathbf{O})}$ (the contour $\Sigma^{(\mathbf{O})}$ is shown in Fig.~\ref{fig:RHP-O}), taking boundary values $\mathbf{O}_+$ ($\mathbf{O}_-$) on each oriented arc of $\Sigma^{(\mathbf{O})}$ from the left (right).  
\item\textbf{Jump Condition}:  The boundary values are related by $\mathbf{O}_+(\lambda)=
\mathbf{O}_-(\lambda)\mathbf{V}^{(\mathbf{O})}$ for $\lambda\in\Sigma^{(\mathbf{O})}$ where the jump matrix $\mathbf{V}^{(\mathbf{O})}$ is as shown in Fig.~\ref{fig:RHP-O}.
\item \textbf{Normalization}:  As $\lambda\to\infty$, $\mathbf{O}(\lambda)\to\mathbb{I}$.
\end{itemize}
The utility of this ``steepest descent'' deformation\footnote{The opening of lenses is traditionally called ``steepest descent'' but the reader will observe that the boundary curves of the lenses are somewhat arbitrary so long as exponential decay is guaranteed by the inequalities on $\phi'(\cdot)$ on the real axis.  Hence one might omit the qualifier ``steepest.''} is now clear:  with the exception of the bands $\bigcup_{n=0}^N[a_n,b_n]$, the jump matrix $\mathbf{V}^{(\mathbf{O})}(\lambda)$ converges to $\mathbb{I}$ pointwise along $\Sigma^{(\mathbf{O})}$ as $\eps\to 0$.
The presence of the bands means that $\mathbf{O}(\lambda)$ is still not the solution of any small-norm RH problem, but we can obtain such a problem via an approximation for $\mathbf{O}(\lambda)$ known as a \emph{parametrix}.

\paragraph{Parametrix construction}
On the band intervals $(a_n,b_n)$ the jump matrix $\mathbf{V}^{(\mathbf{O})}(\lambda)$ has the form
\begin{multline}
\mathbf{V}^{(\mathbf{O})}(\lambda)=\begin{bmatrix}0 & -e^{-2i\phi_n/\eps}\\e^{2i\phi_n/\eps} & 0\end{bmatrix} \\{}+ \text{exponentially small terms as $\eps\to 0$,}
\end{multline}
where $\phi_n$ are well-defined real-valued functions of $(x,t)$ that are independent of $\lambda$ and $\eps$.  We may therefore obtain a formal approximation of $\mathbf{O}(\lambda)$ by solving the following RH problem:  seek $\dot{\mathbf{O}}^{(\mathrm{out})}: \mathbb{C}\setminus\text{bands}\to
\mathrm{SL}(2,\mathbb{C})$ with the properties
\begin{itemize}
\item\textbf{Analyticity}:  $\dot{\mathbf{O}}^{(\mathrm{out})}$ is analytic where defined and takes boundary values $\dot{\mathbf{O}}^{(\mathrm{out})}_\pm(\lambda)$ from $\mathbb{C}_\pm$ on each band $(a_n,b_n)$.
\item\textbf{Jump Condition}:  The boundary values satisfy 
\begin{equation}
\dot{\mathbf{O}}^{(\mathrm{out})}_+(\lambda)=\dot{\mathbf{O}}^{(\mathrm{out})}_-(\lambda)
\begin{bmatrix}0 & -e^{-2i\phi_n/\eps}\\e^{2i\phi_n/\eps} & 0\end{bmatrix},
\end{equation}
where $a_n<\lambda<b_n$, for $n=0,\dots,N$.
\item\textbf{Normalization}:  As $\lambda\to\infty$, $\dot{\mathbf{O}}^{(\mathrm{out})}(\lambda)\to\mathbb{I}$.
\end{itemize}
Since the jump matrix is discontinuous at the band endpoints, we need to specify a singularity at each; we will suppose that for all $n$,
$\dot{\mathbf{O}}^{(\mathrm{out})}(\lambda)=\mathcal{O}((\lambda-a_n)^{-1/4}(\lambda-b_n)^{-1/4})$ as 
$\lambda\to a_n,b_n$.
With this condition, there is a unique solution for $\dot{\mathbf{O}}^{(\mathrm{out})}(\lambda)$
that we call the \emph{outer parametrix}.  In general, it is constructed in terms of Riemann theta functions of genus $N$, but for $N=0$ (one band) the solution is elementary:
\begin{equation}
\dot{\mathbf{O}}^{(\mathrm{out})}(\lambda)=e^{-i\phi_0\sigma_3/\eps}
\mathbf{A}\gamma(\lambda)^{\sigma_3}\mathbf{A}^{-1}
e^{i\phi_0\sigma_3/\eps},\;
\mathbf{A}:=\begin{bmatrix}i & -i\\1 & 1\end{bmatrix},
\end{equation}
where $\gamma(\cdot)$ is the function analytic in $\mathbb{C}\setminus [a_0,b_0]$ that satisfies
\begin{equation}
\gamma(\lambda)^{4}=\frac{\lambda-b_0}{\lambda-a_0}\quad\text{and}\quad
\lim_{\lambda\to\infty}\gamma(\lambda)=1.
\end{equation}

The approximation of the jump matrix $\mathbf{V}^{(\mathbf{O})}$ by piecewise constants, as was used to arrive at the RH problem governing the outer parametrix, is inaccurate near the band endpoints, where the pointwise convergence fails to be uniform.  
In disks
$D_{a_0},\dots,D_{b_N}$ centered at the endpoints, it is therefore necessary to use different (local) approximations of $\mathbf{O}$ called \emph{inner parametrices}.  These are build out of Airy functions. 
We do not give the details here, but a complete description in a setting very similar to the present one can be found in \cite[pgs. 329--334]{MillerQ15}.

\begin{rmk}
We are avoiding the details of the inner parametrices in part to keep the discussion as simple as possible and in part because, provided the match between inner and outer parametrices on the various disk boundaries is a good one, they do not play a role in the leading-order formula for the solution of the defocusing nonlinear Schr\"odinger equation (see \eqref{eq:extract-solution} below).  However there are some problems in which the omitted details are of crucial importance.  For example, when one uses RH methods to describe orthogonal polynomials of large degree \cite{DeiftKMVZ99a,DeiftKMVZ99b}, it is the Airy functions in the inner parametrices that lead to generalizations beyond old results known for classical orthogonal polynomials \cite{Szego75} of Airy asymptotics for the polynomials near turning points and, via the connection with unitary random matrix ensembles, the Airy kernel describing correlation functions of eigenvalues near the edge of the bulk and the concomitant Tracy-Widom law \cite{TracyW94} for the fluctuations of the largest eigenvalue.  This is because to derive these results one must have an explicit approximation for the solution of the corresponding RH problem within analogues of the disks $D_p$.  Another class of examples in which the details of inner parametrices are important includes problems where higher-order corrections to the leading order term are required.  In such problems, the dominant contribution to the error frequently comes from the mismatch between the inner and outer parametrices on the disk boundaries, and calculation of the corrections under suitable scalings of parameters like $x$ and $t$ often leads to universal phenomena.  See \S\ref{sec:universality} for more information.
\end{rmk}

Combining inner and outer parametrices gives rise to an explicit, ad-hoc approximation of $\mathbf{O}(\lambda)$ called the \emph{global parametrix} denoted $\dot{\mathbf{O}}(\lambda)$
and defined piecewise:
\begin{equation}
\dot{\mathbf{O}}(\lambda):=\begin{cases}
\dot{\mathbf{O}}^{(\mathrm{in},p)}(\lambda),&\quad
\lambda\in D_p, \quad p=a_0,\dots,b_N,\\
\dot{\mathbf{O}}^{(\mathrm{out})}(\lambda),&\quad \text{otherwise}.
\end{cases}
\label{eq:piecewise}
\end{equation}
Here $\dot{\mathbf{O}}^{(\mathrm{in},p)}(\lambda)$ denotes the inner parametrix that is installed near $\lambda=p$, which is constructed not only to accurately approximate the jump conditions within $D_p$ but also to accurately match with $\dot{\mathbf{O}}^{(\mathrm{out})}(\lambda)$ on the boundary of $D_p$.  Another important property of the global parametrix guaranteed by the definition \eqref{eq:piecewise} is that $\dot{\mathbf{O}}(\lambda)$ and its inverse are uniformly bounded in the $\lambda$-plane because the outer parametrix is avoided near its singular points where the corresponding inner parametrix is bounded.

\begin{figure}[h]
\begin{center}
\includegraphics[width=\linewidth]{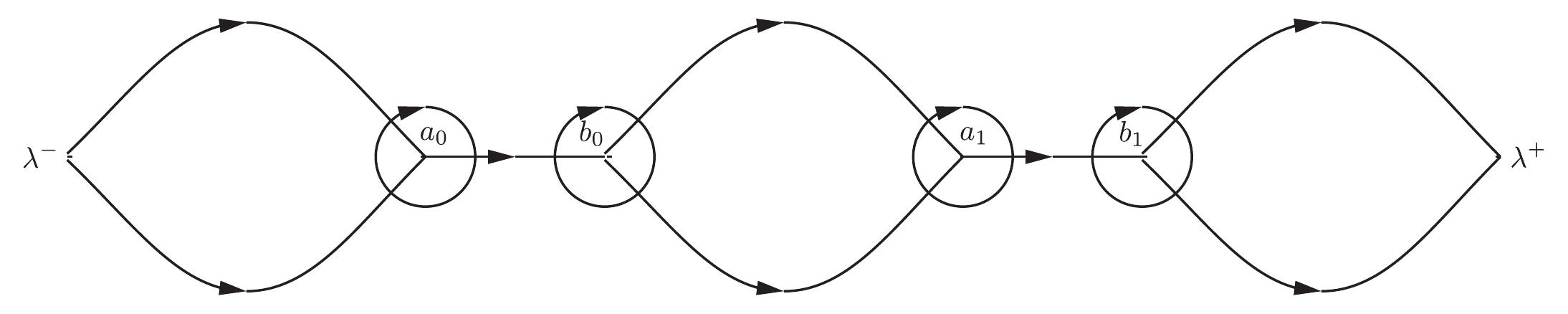}
\end{center}
\caption{The contour $\Sigma^{(\mathbf{E})}$ of the RH problem for the error.}
\label{fig:SigmaE}
\end{figure}

\paragraph{Error analysis by small-norm theory}
Let the \emph{error} of the approximation be defined as the matrix function 
$\mathbf{E}(\lambda):=\mathbf{O}(\lambda)\dot{\mathbf{O}}(\lambda)^{-1}$
wherever both factors make sense.  This makes $\mathbf{E}(\lambda)$ analytic on the complement of an arcwise oriented contour $\Sigma^{(\mathbf{E})}$ (see Fig.~\ref{fig:SigmaE}).
While $\mathbf{O}$ is only specified as the solution of a RH problem, the global parametrix $\dot{\mathbf{O}}(\lambda)$ is known.  Therefore we may regard the mapping $\mathbf{O}\to\mathbf{E}$ as a substitution resulting in an equivalent RH problem for $\mathbf{E}$.

Since both $\mathbf{O}(\lambda)\to\mathbb{I}$ (by normalization condition) and $\dot{\mathbf{O}}(\lambda)\to\mathbb{I}$ (by construction) as $\lambda\to\infty$, we also must have $\mathbf{E}(\lambda)\to\mathbb{I}$ in this limit.
Moreover, by direct calculations, one checks that as a consequence of the uniform boundedness of the outer parametrix outside all disks, 
$\mathbf{E}_+(\lambda)=\mathbf{E}_-(\lambda)(\mathbb{I}+o(1))$ holds as $\eps\to 0$ uniformly for $\lambda\in\Sigma^{(\mathbf{E})}$.
This means that $\mathbf{E}(\lambda)$ satisfies the conditions of a RH problem of small norm type, with estimates of $\mathbf{V}^{(\mathbf{E})}(\lambda)-\mathbb{I}$ in all required spaces being $\bo((\log(\eps^{-1}))^{-1/2})$.  Here, the slow rate of decay as $\eps\to 0$ originates from neighborhoods of the endpoints $\lambda^\pm$, a justification of which can be found in \cite[Lemma 6]{MillerQ15}.
Small-norm theory therefore implies that
$\mathbf{E}(\lambda)$ exists for sufficiently small $\eps$ and is unique, and hence (by unraveling the explicit substitutions) the same is true of $\tilde{\mathbf{M}}(\lambda)$.  Furthermore,
$\mathbf{E}(\lambda)$ has a Laurent series (convergent, because $\Sigma^{(\mathbf{E})}$ is bounded) with moments converging to zero with $\eps$:
\begin{equation}
\mathbf{E}(\lambda)=\mathbb{I}+\sum_{n=1}^\infty \mathbf{E}_n\lambda^{-n}\quad\text{with}\quad
\mathbf{E}_n=o(1),\quad\forall n\ge 1.
\end{equation}

\paragraph{Extraction of the solution}
The last step of the method is to calculate $\tilde{\psi}(x,t)$.  Recall that 
$\mathbf{T}(\lambda)=\mathbb{I}$ and $\dot{\mathbf{O}}(\lambda)=\dot{\mathbf{O}}^{(\mathrm{out})}(\lambda)$ both hold for large enough $|\lambda|$, and that $g(\lambda)\to 0$ as $\lambda\to\infty$.
Therefore, in the semiclassical limit $\eps\to 0$,
\begin{equation}
\begin{split}
\tilde{\psi}(x,t)&=2i\lim_{\lambda\to\infty}\lambda \tilde{M}_{12}(\lambda)\\
&=2i\lim_{\lambda\to\infty}\left[\mathbf{E}(\lambda)\dot{\mathbf{O}}^{(\mathrm{out})}(\lambda)e^{ig(\lambda)\sigma_3/\eps}\right]_{12}\\
&=2iE_{1,12}+2i\dot{O}^{(\mathrm{out})}_{1,12}\\ &=
2i\dot{O}^{(\mathrm{out})}_{1,12}+o(1),
\end{split}
\label{eq:extract-solution}
\end{equation}
where $\dot{\mathbf{O}}^{\mathrm{(out)}}(\lambda)=\mathbb{I}+\dot{\mathbf{O}}^{\mathrm{(out)}}_1\lambda^{-1}+\bo(\lambda^{-2})$ as $\lambda\to\infty$.
When $N=0$ (one band, $(a_0,b_0)$), this reads simply
\begin{equation}
\tilde{\psi}(x,t)=\frac{1}{2}(b_0-a_0)e^{-2i\phi_0/\eps} +o(1),\quad \frac{\partial\phi_0}{\partial x}=\frac{1}{2}(a_0+b_0),
\label{eq:tilde-psi-formula-1}
\end{equation}
where the expression for $\phi_{0,x}$ can be derived by direct construction of the function $g_x$ in terms of Cauchy-type integrals along the lines of what we will describe in \S\ref{sec:construction-of-g}.
In the next simplest case that $N=1$, the Riemann theta functions needed to build $\dot{\mathbf{O}}^{(\mathrm{out})}(\lambda)$ are classical Jacobi theta functions associated with an elliptic curve, and the formula that takes the place of \eqref{eq:tilde-psi-formula-1} instead can be expressed in terms of Jacobi elliptic functions, thus recovering the periodic oscillations of a DSW.

Under some reasonable conditions (in particular these conditions can hold regardless of the number $N$) the $o(1)$ error terms are locally uniform in $(x,t)$.  The leading term $2i\dot{O}^{(\mathrm{out})}_{1,12}$ is a function of $x$, $t$, and $\eps$ that is in general highly oscillatory for small $\eps$.  This shows the strong nature of the semiclassical asymptotics obtained by the DZ steepest descent technique.

\section{Connection with Whitham modulation theory}
\label{sec:ModulationEquations}
\subsection{Construction of $g$}
\label{sec:construction-of-g}
It was shown above that the conditions governing the $g$-function can be seen as the Euler-Lagrange variational conditions for a certain maximization problem.  Conversely, as part of their theory, Lax and Levermore \cite{LaxL83} developed a technique for studying their maximization problem by translating it into a problem of complex analysis involving an analogue of the function $g$.  The question remains:  how does one find the $g$-function in practice?  The relevant techniques for the analysis of the maximization problem of $Q$ defined by \eqref{eq:JinLM-Q} are described in \cite[Section 5]{JinLM99}.  Here, we give the corresponding details of the related maximization problem for the functional $Q$ defined by \eqref{eq:g-function-Q}, i.e., we show how the $g$-function is constructed.  Solving this problem involves, in part, determining the various band and gap subintervals of $(\lambda^-,\lambda^+)$ given $(x,t)\in\mathbb{R}^2$.

The main idea for constructing $g$ is to build a ``candidate'' $\hat{g}$ for $g$ by assuming a certain configuration of bands and gaps and (temporarily) ignoring the inequalities on the functions $\phi$ and $\Delta$.  Later, one uses the inequalities to determine whether the candidate $\hat{g}$ is, in fact, $g$. Suppose that there are $N+1$ bands in $(\lambda^-,\lambda^+)$ that we will denote by $(a_j,b_j)$ with $\lambda^-<a_0<b_0<a_1<b_1<\cdots<a_N<b_N<\lambda^+$.  The complementary intervals are either voids or saturated regions.

Recall that the boundary values of $g$ are subject to:
\begin{itemize}
\item $g_+(\lambda)-g_-(\lambda)=0$ which implies $g'_+(\lambda)-g'_-(\lambda)=0$ for $\lambda$ in voids and outside of $[\lambda_-,\lambda_+]$.
\item $g_+'(\lambda)+g_-'(\lambda)=2\theta'(\lambda)$ for $\lambda$ in bands.
\item $g_+(\lambda)-g_-(\lambda)=2im(\lambda)$ which implies $g_+'(\lambda)-g_-'(\lambda)=2im'(\lambda)$ for $\lambda$ in saturated regions.
\end{itemize}
We therefore know $g_+'-g_-'$ everywhere along $\mathbb{R}$ with the exception of the band intervals, where we know instead $g_+'+g_-'$.  

%
We will now construct a unique candidate $\hat{g}$ that satisfies these conditions, provided that the $2N+2$ band endpoints are appropriately chosen to satisfy an equal number of conditions that will be specified as part of the construction.  The first step is to change the sum of boundary values given in the bands into a difference.  We use a different technique than in the previous section to accomplish this.  Namely, 
consider the function $r(\lambda)$ defined by the relation
\begin{equation}
r(\lambda)^2=\prod_{n=0}^N(\lambda-a_n)(\lambda-b_n)
\label{eq:Riemann-Surface}
\end{equation}
and the additional properties that 
$r(\lambda)$ is analytic for $\lambda\in\mathbb{C}\setminus\bigcup_{n=0}^N
[a_n,b_n]$ and 
$r(\lambda)=\lambda^{N+1}+O(\lambda^N)$ as $\lambda\to\infty$.
Note that the boundary values of $r$ on any band satisfy $r_+(\lambda)+r_-(\lambda)=0$.
Now consider instead of $\hat{g}'(\lambda)$ the function $k(\lambda):=\hat{g}'(\lambda)/r(\lambda)$.  This function is analytic where $\hat{g}'$ is and satisfies
\begin{multline}
k_+(\lambda)-k_-(\lambda)=\\
\begin{cases}
0,&\; \text{$\lambda$ in voids or $\lambda\in\mathbb{R}\setminus[\lambda^-,\lambda^+]$}\\
2\theta'(\lambda)
/
r_+(\lambda)
,&\; \text{$\lambda$ in bands}\\
2im'(\lambda)
/
r(\lambda)
,&\;\text{$\lambda$ in saturated regions}.
\end{cases}
\end{multline}
Up to an entire function (which must be zero for consistency with $\hat{g}'(\lambda)=\bo(\lambda^{-2})$ as $\lambda\to\infty$), $k$ must be given in terms of the difference of its boundary values by the Plemelj formula:
\begin{equation}
k(\lambda)=\frac{1}{\pi i}\int_{\text{bands}}\frac{\theta'(\mu)\,d\mu}{r_+(\mu)(\mu-\lambda)} +
\frac{1}{\pi}\int_{\text{sat'd regs}}\frac{m'(\mu)\,d\mu}{r(\mu)(\mu-\lambda)}.
\end{equation}
We thus obtain the explicit formula $\hat{g}'(\lambda)=r(\lambda)k(\lambda)$.

\begin{figure}[h]
\begin{center}
\includegraphics[width=\linewidth]{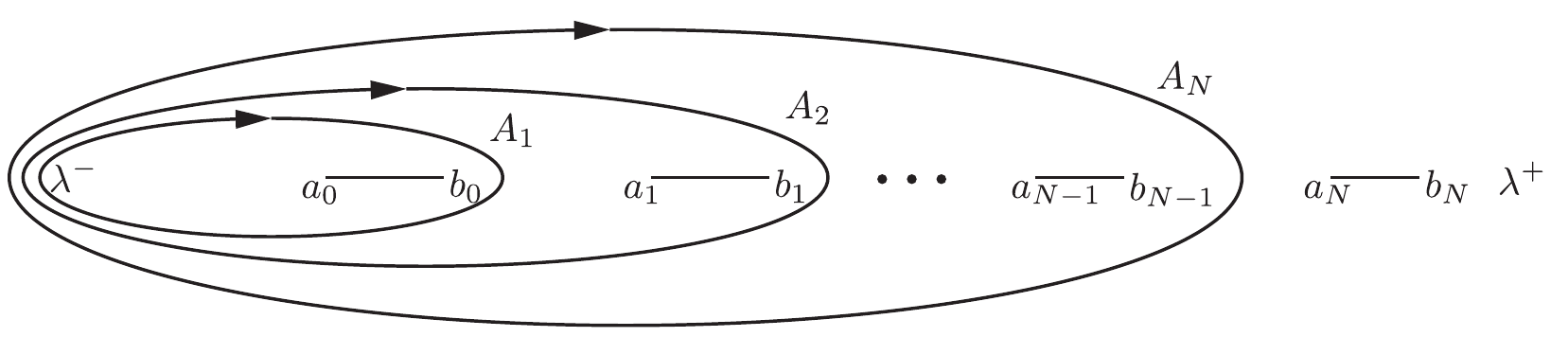}
\end{center}
\caption{The contours $A_j$.}
\label{fig:A-Cycles}
\end{figure}

We have therefore obtained $\hat{g}'(\lambda)$ uniquely given a configuration of bands and gaps, but 
in fact it gives rise to a true candidate $\hat{g}(\lambda)$ only if the endpoints of the bands are appropriately selected, as we now show.  Firstly, we recall that any candidate $\hat{g}$ should satisfy
the additional decay condition $\hat{g}'(\lambda)=\bo(\lambda^{-2})$ as $\lambda\to \infty$; this condition is clearly equivalent to $k(\lambda)=\bo(\lambda^{-(N+3)})$ due to the asymptotic behavior of $r(\lambda)$ for large $\lambda$.  But since $(\mu-\lambda)^{-1}\sim -\lambda^{-1}-\mu\lambda^{-2}-\mu^2\lambda^{-3}+\cdots$, 
$k(\lambda)$ has the Laurent series $k(\lambda)=
-k_1\lambda^{-1}-k_2\lambda^{-2}-k_3\lambda^{-3}-\cdots$
where
\begin{equation}
k_n:=\frac{1}{\pi i}\int_{\text{bands}}\frac{\theta'(\mu)\mu^{n-1}\,d\mu}{r_+(\mu)}+
\frac{1}{\pi}\int_{\text{sat'd regs}}\frac{m'(\mu)\mu^{n-1}\,d\mu}{r(\mu)}.
\end{equation}
Enforcing $k(\lambda)=\mathcal{O}(\lambda^{-(N+3)})$ as $\lambda\to\infty$
requires that 
\begin{equation}
k_n=0\quad \text{for $n=1,\dots,N+2$.}
\label{eq:moments}
\end{equation}
These $N+2$ conditions, which certainly are constraints on the $2N+2$ endpoints of the bands given $(x,t)\in\mathbb{R}^2$, are sometimes called ``moment conditions'' for obvious reasons.  Presuming them to be satisfied, $\hat{g}'(\lambda)$ is integrable at $\lambda=\infty$ so we can obtain $\hat{g}(\lambda)$ vanishing as $\lambda\to\infty$ from $\hat{g}'(\lambda)$ by contour integration:
\begin{equation}
\hat{g}(\lambda)=\int_\infty^\lambda \hat{g}'(\mu)\,d\mu.
\end{equation}
The integration path is arbitrary in the domain $\mathbb{C}\setminus [\lambda^-,\lambda^+]$.
Next observe that while we have arranged that $\hat{g}_+'-\hat{g}_-'=0$ in voids and 
$\hat{g}_+'-\hat{g}_-'=2im'$ in saturated regions, we \emph{actually} require 
$\hat{g}_+-\hat{g}_-=0$ and $\hat{g}_+-\hat{g}_-=2im$  respectively.
Since $m(\lambda^\pm)=0$, one can check that the latter conditions automatically hold in the exterior gaps $(\lambda^-,a_0)$ and $(b_N,\lambda^+)$, however  there remains one ``integral condition'' to impose for each of the $N$ interior gaps.  Let contours\footnote{Half of a homology basis for the hyperelliptic Riemann surface of \eqref{eq:Riemann-Surface}.} $A_1,\dots,A_N$ be as illustrated in Fig.~\ref{fig:A-Cycles}.
Then, the integral conditions are the following.
\begin{itemize}
\item If $(b_n,a_{n+1})$ is a void, then $\hat{g}_+-\hat{g}_-=0$ in this interval is equivalent
to the contour integral condition 
\begin{equation}
\oint_{A_{n+1}} \hat{g}'(\lambda)\,d\lambda=0.
\label{eq:int-cond-void}
\end{equation}
\item If $(b_n,a_{n+1})$ is a saturated region, then $\hat{g}_+-\hat{g}_-=2im$ in this interval is equivalent to the contour integral condition 
\begin{equation}
\oint_{A_{n+1}}\left[\hat{g}'(\lambda)-\frac{1}{\pi}\frac{d}{d\lambda}\int_{\lambda^-}^{\lambda^+}\frac{m(\mu)\,d\mu}{\mu-\lambda}\right]\,d\lambda=0.
\label{eq:int-cond-sat}
\end{equation}
\end{itemize}
Note that in \eqref{eq:int-cond-void}--\eqref{eq:int-cond-sat}, the integrands are analytic on $A_{n+1}$.

If the $2N+2$ moment and integral conditions on the unknowns $a_0,b_0,\dots,a_N,b_N$ have a unique solution, then associated with the symbol sequence $(s_0,s_1,\dots,s_{N+1})$, $s_n=\text{V}$ or $s_n=\text{S}$, indicating the types of gaps in left-to-right order, the candidate $\hat{g}(\lambda)$ is uniquely determined by integration and it satisfies all desired conditions of a $g$-function except possibly the inequalities on the corresponding functions $\hat{\phi}$ and $\hat{\Delta}$.  The expectation is that enforcing these inequalities on the candidate $\hat{g}$ (thereby making $\hat{g}=g$) should select both the value of $N$ (i.e., the genus of the Riemann surface of the equation \eqref{eq:Riemann-Surface}) and the symbol sequence $(s_0,\dots,s_{N+1})$.

\begin{figure}[h]
\begin{center}
\includegraphics[width=\linewidth]{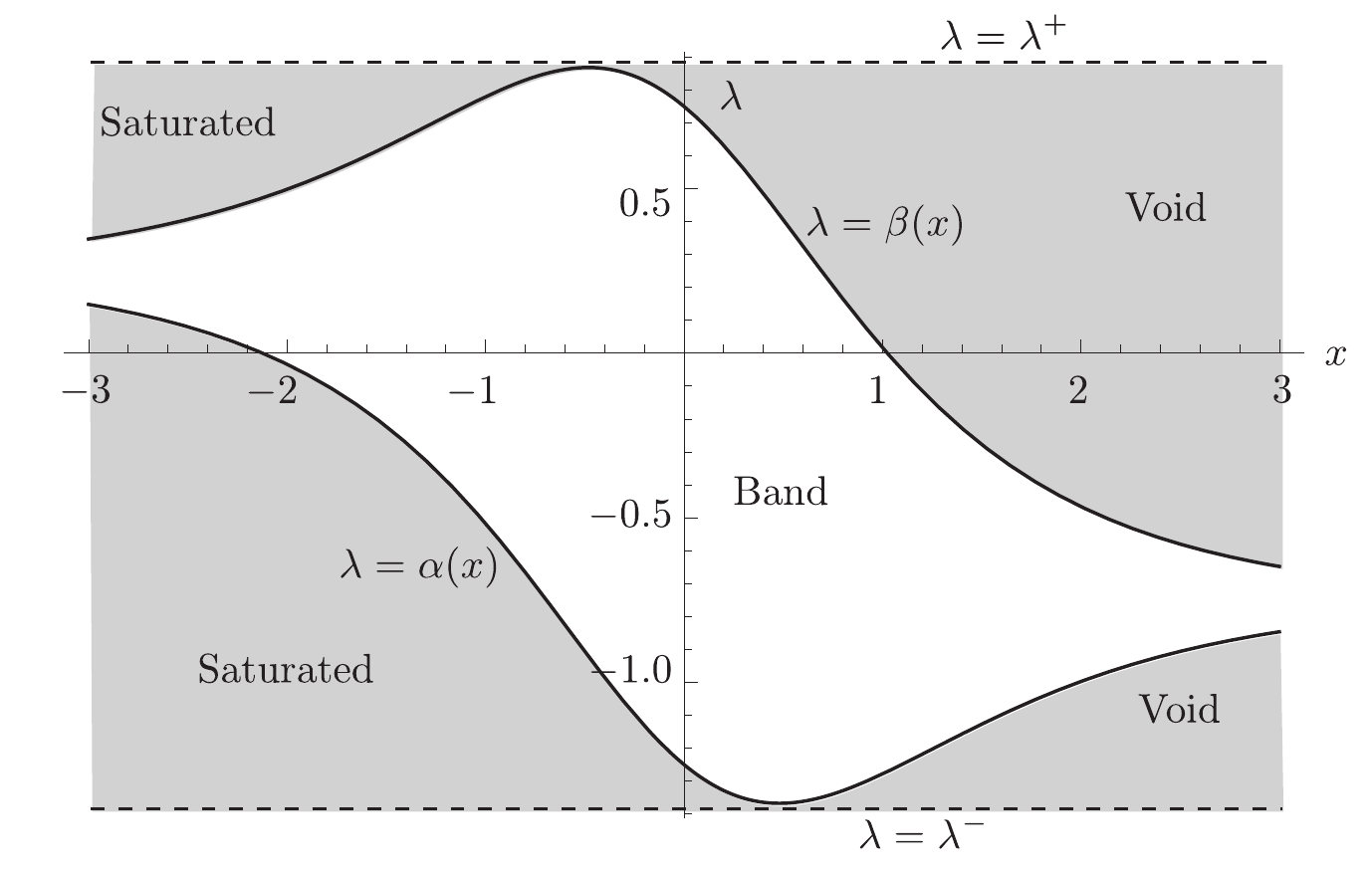}
\end{center}
\caption{The voids, bands, and saturated regions as functions of $x$ when $t=0$.}
\label{fig:VandS}
\end{figure}

The procedure in practice is therefore to determine $N$ and $(s_0,\dots,s_{N+1})$ so that the inequalities hold.  The independent variables $x$ and $t$ are parameters in this procedure.  In particular, $N=N(x,t)$.
To give some further details of this procedure in the simplest case,  we first claim that the $g$-function can be determined explicitly when $t=0$, and that $N=0$ (one band) suffices in this case.
Recall that for $N=0$ there are just two conditions to be satisfied by the endpoints $a_0,b_0$:  $k_1=k_2=0$.  We have the following result.
\begin{prop}
Set $t=0$.  The equations $k_1=k_2=0$ are simultaneously satisfied by 
\begin{equation}
a_0=\alpha(x)\quad\text{and}\quad b_0=\beta(x)
\end{equation}
with symbol sequences
$(\mathrm{V},\mathrm{V})$ where $\alpha'(x)>0$ and $\beta'(x)<0$,
$(\mathrm{V},\mathrm{S})$ where $\alpha'(x)>0$ and $\beta'(x)>0$,
$(\mathrm{S},\mathrm{V})$ where $\alpha'(x)<0$ and $\beta'(x)<0$, and
$(\mathrm{S},\mathrm{S})$ where $\alpha'(x)<0$ and $\beta'(x)>0$.
\label{prop:t-zero}
\end{prop}
One can further confirm that the necessary inequalities are also satisfied by the specified configuration when $t=0$.  This information is summarized in Fig.~\ref{fig:VandS}.
Note that when $t=0$, void intervals (resp. saturated regions) do not contain the point $\lambda=-\tfrac{1}{2}u_0(+\infty)$ (resp., $\lambda=-\tfrac{1}{2}u_0(-\infty)$) for any $x\in\mathbb{R}$.  For analytic initial data, this implies that the function $\phi$ is analytic in all gaps when $t=0$.
Combining Proposition~\ref{prop:t-zero} with the formula \eqref{eq:tilde-psi-formula-1} for $\tilde{\psi}(x,t)$ essentially completes the proof by the DZ steepest descent technique that the difference $\tilde{\psi}(x,0)-\psi(x,0)$ between the initial condition corresponding to the modified scattering data and the true initial condition\footnote{The step of modification of the initial data is very common in the literature, going back to Lax and Levermore \cite{LaxL83}.  It can be avoided, i.e., the initial-value problem can be directly studied in the limit $\eps\to 0$, provided that one has sufficiently accurate WKB-type asymptotics of the scattering data, including approximations with error estimates valid near transitional points in the spectrum. This kind of information has become available fairly recently and only for the Schr\"odinger operator with repulsive potential \cite{Ramond96}.  Using the results of \cite{Ramond96},
Claeys and Grava \cite{ClaeysG09,ClaeysG10a,ClaeysG10b} have completely avoided ad-hoc approximations of the scattering data in their analysis of the zero-dispersion limit for KdV.} converges to zero with $\epsilon$ locally uniformly in $x$.

\begin{figure}[h]
\begin{center}
\includegraphics[width=0.8\linewidth]{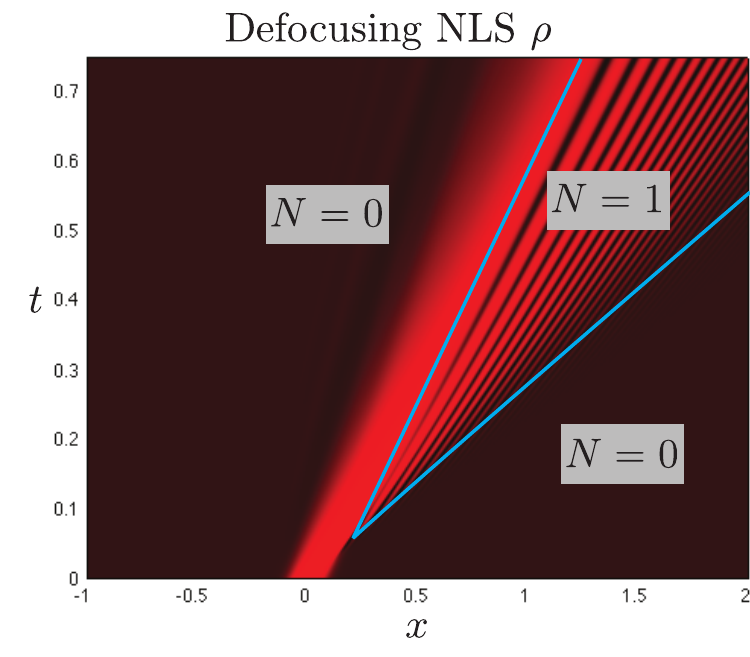}
\end{center}
\caption{From \cite{DifrancoMM11}.  The density $\rho(x,t)$ of a numerical solution of the defocusing NLS equation with initial data $\rho_0(x)=\tfrac{1}{10}+\tfrac{1}{2}e^{-256x^2}$, 
$u_0(x)=1$, and with
$\eps=0.0122$.
(Numerically periodic boundary conditions enforced.)}
\label{fig:DefocNLSBreak}
\end{figure}

The implicit function theorem can be used to continue the solution to $k_1=k_2=0$ for small $t$ independent of $\eps$.  The necessary inequalities also persist as they hold strictly when $t=0$. 
Therefore we have a genus $N=0$ configuration of a single band for all $x\in\mathbb{R}$ if $t$ is sufficiently small.  For larger $t$, 
it becomes necessary for some $x\in\mathbb{R}$ to pass to a larger genus $N$ because either the solution of $k_1=k_2=0$ cannot be continued or an inequality is violated.
The earliest point of transition is the shock time for the dispersionless NLS system, after which one observes a DSW corresponding to genus $N=1$
in a wedge of the $(x,t)$-plane anchored at the breaking point.  See Fig.~\ref{fig:DefocNLSBreak}.
We therefore observe the following key point:  \emph{genus bifurcations in the $g$-function are the integrable nonlinear analogues of stationary phase point bifurcations in the linear theory}.  They are ``nonlinear caustics'', and they form the boundaries of various types of DSWs in the $(x,t)$-plane.

\subsection{Modulation equations}
To explain the connection with Whitham modulation equations, first recall the general fact first explained in \cite{FlaschkaFM80} that the (multiphase) Whitham modulation equations for integrable systems
can always be represented in Riemann invariant form.  To see where these equations arise in the present context, consider a domain in the $(x,t)$-plane in which the genus $N=N(x,t)$ is constant.  We claim that
the band endpoints $a_0,b_0,\dots,a_N,b_N$, which depend on $(x,t)$ in the selected domain, are exactly the Riemann invariant variables for the Whitham equations.  Indeed, these functions are necessarily solutions of the $N+2$ moment conditions and $N$ integral conditions, which we combine into $2N+2$ algebraic (i.e., not differential) equations of the general form $e_n(a_0,b_0,\dots,a_N,b_N;x,t)=0$.  Each of the functions $e_n$ is linear in $x$ and $t$, but highly nonlinear in $a_0,\dots,b_N$, and the functions $e_n$ encode the initial data.  Since the band endpoints depend on $(x,t)$ via these equations,
implicit differentiation yields the following:
\begin{equation}
\sum_{j=0}^N\left[\frac{\partial e_n}{\partial a_j}\frac{\partial a_j}{\partial x,t}+\frac{\partial e_n}{\partial b_j}\frac{\partial b_j}{\partial x,t}\right]=-\frac{\partial e_n}{\partial x,t},\quad n=1,\dots,2N+2.
\end{equation}
The right-hand side has no explicit dependence on $(x,t)$ by linearity of $e_n$ in these variables, and when evaluated on a solution of the equations $e_n=0$, $n=1,\dots,2N+2$, the coefficient matrix of the partial derivatives of the band endpoints can also be written in terms of the endpoints alone\footnote{We omit these calculations, but see \cite[Sec.\@ 4.4]{BuckinghamM13}, \cite[Sec.\@ 5.3]{KamvissisMM03}, or \cite[Sec.\@ 4.3]{DifrancoM13} for more details in some examples.}.  Whenever this coefficient matrix is invertible, one therefore obtains the partial derivatives of the band endpoints with respect to $(x,t)$ in terms of the band endpoints themselves.  It follows that there exist functions $c_j$ and $d_j$ of the band endpoints, $j=0,\dots,N$, such that
\begin{equation}
\frac{\partial a_j}{\partial t}+c_j(a_0,\dots,b_N)\frac{\partial a_j}{\partial t}=0,\;
\frac{\partial b_j}{\partial t}+d_j(a_0,\dots,b_N)\frac{\partial b_j}{\partial t}=0.
\end{equation}
These are precisely the $N$-phase Whitham equations for the defocusing NLS equation rendered in Riemann-invariant form.  For example,
in the special case of $N=0$,
implicit differentiation of the conditions $k_1=0$ and $k_2=0$ with respect to $x$ and $t$ shows that
the following equations hold true:
\begin{equation}
\frac{\partial a_0}{\partial t}-\left[\frac{3}{2}a_0+\frac{1}{2}b_0\right]\frac{\partial a_0}{\partial x}=0,\;
\frac{\partial b_0}{\partial t}-\left[\frac{3}{2}b_0+\frac{1}{2}a_0\right]\frac{\partial b_0}{\partial x}=0.
\end{equation}
Setting $a_0=-\tfrac{1}{2}u-\sqrt{\rho}$ and $b_0=-\tfrac{1}{2}u+\sqrt{\rho}$, this system becomes the dispersionless defocusing NLS system, i.e., \eqref{eq:Madelung-general} with $\sigma=1$ and $\eps$ set to zero.


These considerations show that in each region of the $(x,t)$-plane corresponding to fixed $N$, the DZ method approximates the solution $\tilde{\psi}(x,t)$ as a modulated $N$-phase wave whose microscopic oscillations are described via the outer parametrix $\dot{\mathbf{O}}^{(\mathrm{out})}(\lambda)$ that is given in general in terms of the theta functions associated with the genus $N$ Riemann surface associated with the equation \eqref{eq:Riemann-Surface}, and whose macroscopic deformations entering via the (slow) $(x,t)$-dependence of the band endpoints are governed by the multiphase form of Whitham's modulation theory.  It was observed already in the older context of the LL method \cite{JinLM99,LaxL83} that the endpoints of intervals in which the maximizer $\eta$ achieves either its upper or lower constraints are Riemann invariants for the Whitham modulation equations, but the weak nature of the limit obtained precludes the method from directly exhibiting the modulated wavetrain that is established by the DZ method.  Revealing the oscillations via the LL method requires a higher-order theory due to Venakides \cite{Venakides90} that is challenging to make completely rigorous.


\subsection{Global analysis for problems with elliptic Whitham PDEs}
While one may derive the Whitham modulation equations from the moment and integral conditions on the band endpoints by implicit differentiation, another key point is that the latter conditions are an implicit algebraic representation of the \emph{solutions} of the Whitham PDEs that are actually relevant to the analysis of the function $\tilde{\psi}(x,t)$ in various parts of the $(x,t)$-plane.  These conditions are thus related to the generalized hodograph method of Tsar\"ev \cite{Tsarev85}, with the important point that the arbitrary functions involved are explicitly connected to initial data.  

From the point of view of the DZ method and the conditions on the $g$-function, the fact that the band endpoints satisfy certain quasilinear partial differential equations in Riemann-invariant form therefore appears to be rather auxiliary.  This further suggests that analytical issues associated with formulating Cauchy problems for the Whitham PDEs may not be so important after all.  The most challenging problems\footnote{Whitham called such problems \emph{modulationally unstable}.} from the latter perspective are those for which the Whitham equations are a quasilinear system of elliptic type, as arise in the modulation theory of the focusing NLS equation (\eqref{eq:NLS-general} with $\sigma=-1$) or the sine-Gordon equation (\eqref{eq:SG} below).  Cauchy problems for such systems are ill-posed.  Indeed, with the method of characteristics unavailable, the only way to solve a Cauchy problem for an elliptic system is to appeal to the Cauchy-Kovaleskaya method, which requires analyticity of the given data, perhaps a physically unreasonable mathematical abstraction.  However,
the DZ method has been applied to such systems \cite{BuckinghamM13,KamvissisMM03,TovbisVZ04}, generating solutions of the corresponding elliptic Whitham equations in implicit form, and proving that certain initial-value problems for integrable equations with elliptic Whitham equations indeed have approximate solutions in the semiclassical limit that are governed by the latter elliptic equations.  Analyticity of the initial data enters into these analyses,  not via the direct construction of solutions of elliptic Cauchy problems, but rather via the need to analytically continue certain functions like $m$ and $\theta_0$ into the complex plane to capture bands and gaps that have left the real axis with varying $(x,t)$.
\section{Universality}
\label{sec:universality}
\subsection{The concept of universality in mathematical physics}
Consider a piece of iron placed in a magnetic field.  Provided the temperature $T$ is below a certain \emph{critical temperature} $T_\mathrm{c}$, the sample exhibits a residual magnetization $M_0(T)>0$ after the externally imposed field has been slowly turned off.  The critical temperature is a point of \emph{phase transition} for the system.  If we consider the asymptotic behavior of the function $M_0(T)$ in the vicinity of the critical temperature, a power law is observed involving a \emph{critical exponent} $\beta>0$:
\[
M_0(T)\sim [-t]_+^\beta,\quad t:=\frac{T-T_\mathrm{c}}{\Delta T},\quad T\to T_\mathrm{c},
\]
where $\Delta T$ is a certain positive constant and $[\cdot]_+$ denotes the positive part.  The simplest statement of \emph{universality} in mathematical physics is that many different physical systems have the same critical exponents.  

\begin{figure}[h]
\begin{center}
\includegraphics[width=\linewidth]{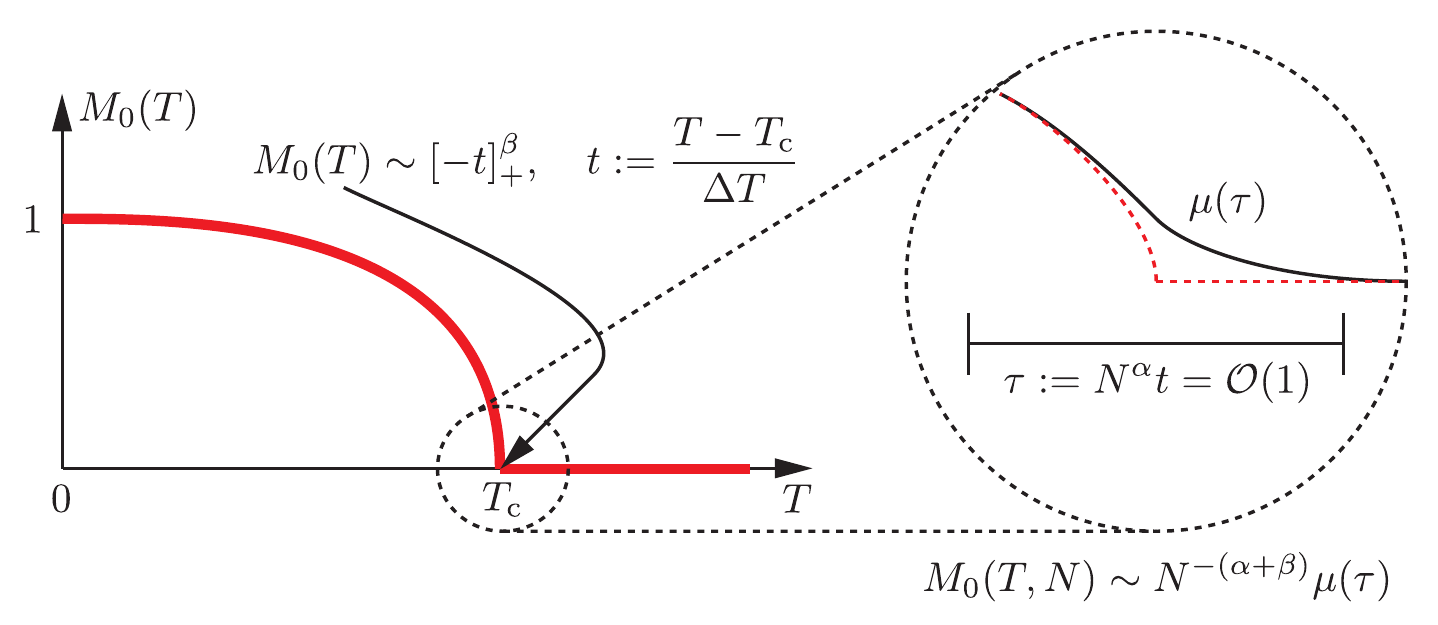}
\end{center}
\caption{Residual magnetization as a function of temperature.}
\label{fig:ResidualMagnetization}
\end{figure}

But as a property of a macroscopic piece of iron, the residual magnetization $M_0(T)$ actually arises from another quantity, $M_0(T,N)$ describing the residual magnetization of a system consisting of $N$ particles, in the \emph{thermodynamic limit} $N\to\infty$.  That is, for each $T>0$, $M_0(T,N)\to M_0(T)$ as $N\to\infty$.  Since the critical temperature $T_\mathrm{c}$ is a point of non-smoothness of the limiting function $M_0(T)$, it is interesting to consider the so-called \emph{double-scaling limit} of $M_0(T,N)$, in which $N\to\infty$ while simultaneously $T\to T_\mathrm{c}$ at a suitable rate.  Indeed,
for a special choice of an exponent $\alpha>0$, we may set $\tau:=N^\alpha t$ and, now holding $\tau\in\mathbb{R}$ fixed, consider the limit
of $N^{\alpha+\beta}M_0(T,N)$ as $N\to\infty$.  For the correct value of $\alpha$ we find a smooth limiting function $\mu(\tau)$, which serves to locally regularize the behavior of $M_0$ near the critical temperature.  See Fig.~\ref{fig:ResidualMagnetization} for a qualitative sketch.  
The function $\mu(\cdot)$ is a feature of the system like the critical exponent $\beta$, and a more sophisticated notion of universality is that many different physical systems reveal the same function $\mu(\cdot)$ in an appropriate double-scaling limit.  A similar phenomenon is apparent in the central limit theorem of probability, where $\mu$ is identified with the standard normal distribution.

\subsection{Universality for weakly-dispersive linear waves}
We may regard points on the caustic curves separating regions of the $(x,t)$-plane containing oscillations described by modulated Riemann theta functions of different genera $N$ (or, in the linear case, superpositions of different numbers of waves) as analogues of phase transition points, and hence we may examine the properties of solutions near these points to develop the idea of universality in wave propagation problems.  As a first example, we return to the linear Schr\"odinger equation, namely \eqref{eq:NLS-general} with $\sigma=0$ and initial data \eqref{eq:schrod-initial-data}.

The solution $\psi(x,t)$ exhibits transitional behavior near the caustic curves $x=x^\pm(t)$ described in \S\ref{sec:linear}, and the
stationary phase formula \eqref{eq:stationary-phase-formula} fails to be accurate near these curves.
To analyze the solution near a caustic,
fix $t_0>t_c$ and consider $x$ near $x_0:=x^-(t_0)$.  As  $x\to x_0$, the two most positive critical points $y_1<y_2$ of $I(y;x,t_0)$ coalesce at some point $\tilde{y}_{12}$ while the most negative critical point tends to a distinct value: $y_0\to \tilde{y}_0$.  This coalescence is what makes the classical method of stationary phase break down.
However, in 1957, Chester, Friedman, and Ursell \cite{ChesterFU57} developed a generalization of the stationary phase method applying to such degenerate situations.  When $y\approx \tilde{y}_{12}$ the phase
behaves like a cubic, and the key idea in \cite{ChesterFU57} is to try to write the equation
$I(y;x,t_0)=\tfrac{1}{3}z^3-b(x)z + c(x)$ by choosing appropriate $b(x)$ and $c(x)$ with $b(x_0)=0$ and $c(x_0)=I(\tilde{y}_{12};x_0,t_0)$.
This has the effect of making the phase \emph{exactly} cubic in terms of $z=z(y)$ with $z(\tilde{y}_{12})=0$.  It is nontrivial but possible to find
nice functions $b=b(x)$ and $c=c(x)$ so that $z=z(y)$ exists as a univalent conformal map.  Then
\begin{multline}
\int_{\tilde{y}_{12}-\delta}^{\tilde{y}_{12}+\delta}e^{iI(y;x,t)/\eps}\sqrt{\rho_0(y)}\,dy \\
=\eps^{1/3}e^{ic(x)/\eps}\int_{z(\tilde{y}_{12}-\delta)/\eps^{1/3}}^{z(\tilde{y}_{12}+\delta)/\eps^{1/3}}
e^{i(w^3/3+\xi w)}f(\eps^{1/3}w)\,dw,
\end{multline}
where $f(z):=\sqrt{\rho_0(y(z))}y'(z)$, $\xi:=-b(x)/\eps^{2/3}$, and $w=\eps^{-1/3}z(y)$.
Approximating the slowly-varying factor $f(\eps^{1/3}w)$ by $f(0)$ and letting $\eps\downarrow 0$ yields the well-known integral representation of the Airy function \cite{dlmf}:
\begin{equation}
\mathrm{Ai}(\xi):=\lim_{R\uparrow\infty}\frac{1}{2\pi}\int_{-R}^R e^{i(w^3/3+\xi w)}\,dw. 
\end{equation}
%
The result of the generalized stationary phase method \cite{ChesterFU57} is therefore the following formula:
\begin{multline}
\psi(x,t_0)=
\frac{e^{-i\pi/4}}{\eps^{1/6}}e^{i\varphi_{12}/\eps}e^{i(x-x_0)U_{12}/\eps}L_{12}
\mathrm{Ai}\left(-M_{12}\frac{x-x_0}{\eps^{2/3}}
\right) \\
{}+ e^{i\varphi_0/\eps}e^{i(x-x_0)U_0/\eps}N_0+\bo(\eps^{1/6})
\label{eq:CFU-formula}
\end{multline}
valid as $\eps\to 0$ for $x-x_0=\bo(\eps^{2/3})$, where
\begin{equation}
\varphi_{12}:=I(\tilde{y}_{12};x_0,t_0),\; U_{12}:=u_0(\tilde{y}_{12}),
\; K_{12}:=\left[\frac{2}{I'''(\tilde{y}_{12};x_0,t_0)}\right]^{1/3}
\end{equation}
are basic constants associated with the degeneration point $\tilde{y}_{12}$ from which we also obtain
$L_{12}:=\sqrt{2\pi\rho_0(\tilde{y}_{12})}K_{12}/t_0^{1/2}$ and $M_{12}:=K_{12}/t_0$,
and where
\begin{equation}
\varphi_0:=I(\tilde{y}_0;x_0,t_0),\; U_0:=u_0(\tilde{y}_0),\; N_0:=\sqrt{\frac{\rho_0(\tilde{y}_0)}{t_0I''(\tilde{y}_0;x_0,t_0)}}
\end{equation}
are basic constants associated with the limiting simple stationary phase point $\tilde{y}_0$.  The formula \eqref{eq:CFU-formula} is valid at an arbitrary smooth point on the caustic curve; however when $(x,t)$ is in a neighborhood of the breaking point $x_c=x^\pm(t_c)$, all three stationary phase points merge and a different approximation based on the Pearcey integral (quartic exponent) in place of the Airy integral (cubic exponent) is required.

The result \eqref{eq:CFU-formula} shows that for generic points on the caustic curve for the linear Schr\"odinger equation, the classical Airy function $\mathrm{Ai}(\cdot)$ is an analogue of the function $\mu(\cdot)$; it describes the asymptotic behavior of the solution $\psi(x,t)$
near the caustic, \emph{regardless of the details of the initial data that generated the caustic in the first place}.  In this way, the Airy function describes universality for the linear Schr\"odinger equation.  More remarkably, this type of universality is valid not only for arbitrary initial data, but also for arbitrary linear dispersive wave equations.  Indeed, the reader will observe that the only true hypothesis needed for the above arguments is the degeneration of two simple critical points of the phase in a Fourier-type integral.  

\subsection{Nonlinear examples}
In a series of remarkable papers, Claeys and Grava have applied the DZ method to the RH problem for the KdV equation \eqref{eq:KdV}
with negative Schwartz-class initial data $u_0$ having a single minimum point.  In the small-dispersion limit $\eps\to 0$, $u(x,t)$ is at first approximated by the corresponding solution of the IB equation $u_t+2uu_x=0$, which breaks down at a time $t_c>0$ and a breaking point $x=x_c$. The methodology of \cite{DeiftVZ97} shows that there are curves $x^-(t)<x^+(t)$ for $t>t_c$ with $x^-(t_c)=x^+(t_c)=x_c$ between which there exists a DSW described by modulated elliptic functions ($N=1$) matching onto IB solutions ($N=0$) outside the interval $(x^-(t),x^+(t))$.
Claeys and Grava both made the techniques of \cite{DeiftVZ97} completely rigorous by including error estimates for the approximation of the reflection coefficient from \cite{Ramond96} and by providing their own error estimates for the inverse problem.  But rather than considering asymptotics for fixed $(x,t)\in\mathbb{R}^2$, they considered various double-scaling limits associated with neighborhoods of the caustic curves $x^\pm(t)$.
One of their results \cite{ClaeysG10b} is a natural nonlinear analogue of the Airy function universality for linear dispersive waves, namely that near the curve $x=x^-(t)$, the leading edge of the DSW, an asymptotic approximation similar to \eqref{eq:CFU-formula} holds for $u(x,t)$, but in which the universal profile function is no longer the Airy function $\mathrm{Ai}(\cdot)$ but is instead the \emph{Hastings-McLeod solution} $w=w_\mathrm{HM}(z)$ of the homogeneous \emph{Painlev\'e-II equation} $w''(z)=zw(z)+2w(z)^3$ uniquely specified by the boundary condition that $w_\mathrm{HM}(z)\sim \mathrm{Ai}(z)$ as $z\to +\infty$.  The Painlev\'e-II equation \cite{dlmf} is obviously a nonlinear generalization of the Airy equation $w''(z)=zw(z)$ satisfied by $w=\mathrm{Ai}(z)$.  Another result \cite{ClaeysG10a} concerns the asymptotic behavior of $u(x,t)$ near the trailing edge of the DSW, $x\approx x^+(t)$.  Here again a universal wave profile emerges from a double-scaling limit combined with the DZ method, with the profile consisting of a train of KdV solitons whose precise locations are determined remarkably from the normalization constants of the famous Hermite polynomials.  

Perhaps the most impressive result of Claeys and Grava on this topic is their proof of universality of a KdV analogue of the Pearcey integral.
\begin{thm}[Claeys \& Grava \cite{ClaeysG09}]
Let $(u_c,x_c,t_c)$ denote the first shock of the IB equation $u_t+2uu_x=0$ with $u(x,0)=u_0(x)$.
If, with $k:=-(u_0^{-1})_-'''(u_c)$, the rescaled independent variables
\begin{equation}
X=\frac{x-x_c-2u_c(t-t_c)}{(8k\eps^6)^{1/7}}\quad\text{and}\quad T=\frac{2(t-t_c)}{(4k^3\eps^4)^{1/7}}
\end{equation}
are held fixed while $\eps\to 0$, then for a very general class of $u_0$, 
\begin{equation}
u(x,t;\eps)=u_c +\left(\frac{2\eps^2}{k^2}\right)^{1/7}U(X,T)+\bo(\eps^{4/7}),\quad \eps\to 0,
\end{equation}
where $U(X,T)$ is the unique smooth real solution of the second Painlev\'e-I equation:
$
X=TU-[\tfrac{1}{6}U^3+\tfrac{1}{24}(U_X^2+2UU_{XX})+\tfrac{1}{240}U_{XXXX}]
$.
\label{thm:ClaeysG09}
\end{thm}

\begin{figure}[h]
\begin{center}
\includegraphics[width=\linewidth]{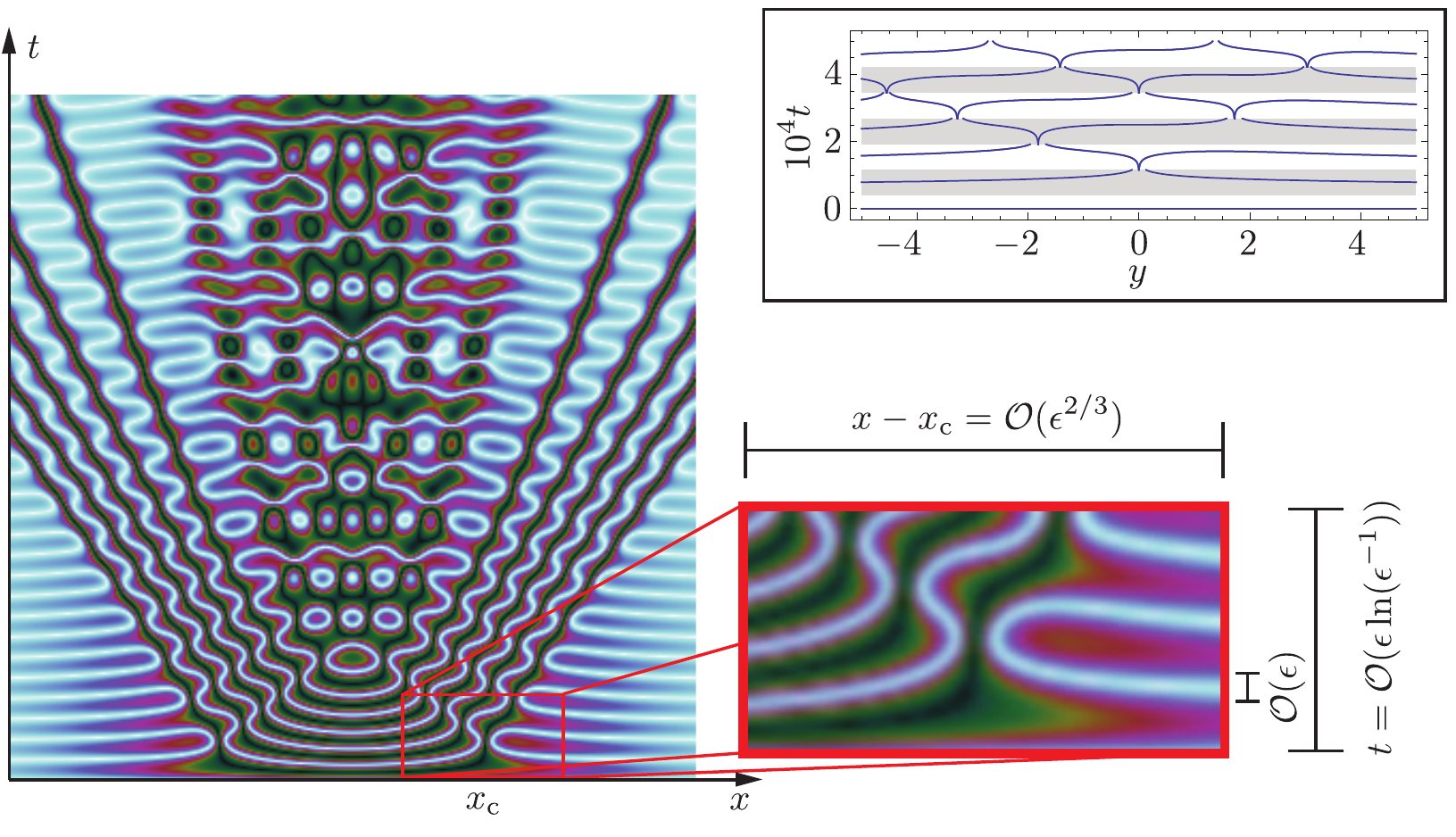}
\end{center}
\caption{Left-hand plot:  a density plot of $\cos(u(x,t))$ (lighter colors for $\cos(u)\approx 1$ and darker colors for $\cos(u)\approx -1$) for $0<t<5$ and $-2.5<x<2.5$ for pure impulse initial data with $G(x)=-3\mathrm{sech}(x)$ and $\eps=0.046875$.  Lower right-hand inset: a blow-up of the region near the critical point $(x,t)=(x_c,0)$.  Upper right-hand inset:  the theoretical locations $T=0$ of the kinks for $n=0,1,\dots,6$ superimposed on the alternately shaded strips $S_0,S_1,\dots,S_6$ for $\eps=10^{-5}$ and $4\nu^{1/3}=1$.}
\label{fig:SG-zoom}
\end{figure}

The Painlev\'e transcendents \cite{dlmf} often appear as universal wave profiles.  For singular limits involving integrable dispersive equations, this is not a surprise because the Painlev\'e equations are themselves integrable systems with corresponding RH problems characterizing their solutions.  The latter RH problems play the role of inner parametrices in the DZ analysis.  Another example of universality for nonlinear waves in which functions satisfying Painlev\'e equations arise is in the analysis of solutions of the semiclassical sine-Gordon equation 
\begin{equation}
\eps^2 u_{tt}-\eps^2 u_{xx}+\sin(u)=0
\label{eq:SG}
\end{equation}
subject to ``pure impulse'' initial data $u(x,0)=0$, $\eps u_t(x,0)=G(x)$, where $G(x)$ is a negative bell-shaped Schwartz-class analytic function with 
a minimum value $\min G(x)<-2$.  Letting $G$ be even for simplicity, there are two critical points $\pm x_c$, $x_c>0$ defined by $G(x_c)=-2$, such that:
for $|x|<x_c$ and $t$ small but independent of $\eps$, $u(x,t)$ is approximated by a modulated superluminal kink train subject to hyperbolic Whitham equations, while for $|x|>x_c$ and $t$ small, $u(x,t)$ is approximated by a modulated train of low-amplitude waves subject instead to elliptic Whitham equations \cite{BuckinghamM13}.  In a double-scaling limit chosen to zoom in on the point $(x,t)=(x_c,0)$, the following result is proved in \cite{BuckinghamM12}.  Here, $\tilde{u}(x,t)$ is a reflectionless approximation to $u(x,t)$ for which $\tilde{u}(x,0)\to 0$ and $\eps\tilde{u}_t(x,0)\to G(x)$, locally uniformly in $x$, as $\eps\to 0$.
\begin{thm}
Suppose $\nu:=[12G'(x_c)]^{-1}>0$.
Fix an integer $n$ and assume that $(x,t)$ lies in the horizontal strip $S_n$ in the $(x,t)$-plane
given by the inequality
$|t-\tfrac{2}{3}n\eps\ln(\eps^{-1})|\le\tfrac{1}{3}\eps\ln(\eps^{-1})$.
Suppose also that 
$x-x_c = \bo(\eps^{2/3})$.  Then
\begin{equation}
\begin{split}
\cos(\tfrac{1}{2}\tilde{u}(x,t))&=(-1)^n \mathrm{sgn}(\mathcal{U}_n(y))\,\mathrm{sech}(T) + 
E_{\cos}(x,t)
\\
\sin(\tfrac{1}{2}\tilde{u}(x,t))&=(-1)^{n+1}\tanh(T)+
E_{\sin}(x,t)
\end{split}
\label{eq:SG-asympt-formula}
\end{equation}
where $E_{\cos}$ and $E_{\sin}$ are error terms vanishing\footnote{except near  distinguished points associated with singularities of $\ln(|\mathcal{U}_n|)$.} as $\eps\to 0$,  and
\begin{equation}
T:=\frac{t}{\eps}-2n\ln\left(\frac{4\nu^{1/3}}{\eps^{1/3}}\right) +
\ln(|\mathcal{U}_n(y)|),\;
y:=\frac{x-x_c}{2\nu^{1/3}\eps^{2/3}}.
\end{equation}
\end{thm}
The functions $\mathcal{U}_n(y)$ are rational functions of $y$ generated (for $n>0$) from $\mathcal{U}_0(y):=1$ by the explicit recursion
$\mathcal{U}_{n+1}(y):=-\tfrac{1}{6}y\mathcal{U}_n(y)-\mathcal{U}_n'(y)^2/\mathcal{U}_n(y)+\tfrac{1}{2}\mathcal{U}_n''(y)$.  Although they are in a sense elementary functions, they are also Painlev\'e solutions in that $w(y):=\mathcal{U}_n'(y)/\mathcal{U}_n(y)$ is the unique rational solution of the inhomogeneous Painlev\'e-II equation $w''(y)=\tfrac{2}{3}yw(y)+2w(y)^3-\tfrac{2}{3}n$.  The asymptotic formula \eqref{eq:SG-asympt-formula} shows that $\tilde{u}(x,t)$ is approximated in the strip $S_n$ by an isolated kink that is centered along a spacelike curve $T=0$; the main role of the function $\mathcal{U}_n(y)$ in the formula is to identify the curve $T=0$ as a translate of the graph of $t=-\eps\ln(|\mathcal{U}_n(y)|)$.  Thus, the solution $\tilde{u}(x,t)$ resembles a train of kinks, but the kinks are located according to the graphs of the Painlev\'e-II rational functions.  See Fig.~\ref{fig:SG-zoom}.
Universality near wave caustics has also been investigated in other nonlinear systems having elliptic Whitham modulation equations, for example the focusing NLS equation \cite{BertolaT10,BertolaT13}.

\subsection{Broader universality conjectures}
The appearance of the particular solution $U(X,T)$ of the second Painlev\'e-I equation (with parameter $T\in\mathbb{R}$) in Theorem~\ref{thm:ClaeysG09} was predicted several years earlier by Dubrovin \cite{Dubrovin06}, who developed a careful but formal asymptotic analysis of a wide class of weakly dispersive perturbations of the IB equation $u_t+2uu_x=0$ under the assumption that $x$ and $t$ were suitably localized near the \emph{gradient catastrophe} (shock) point $(x_c,t_c)$ for the unperturbed problem.  The goal of this analysis was to determine the dispersive wave field near $(x_c,t_c)$ and how it depends on initial data and the nature of the dispersive corrections.  Dubrovin argued that
the function $U(X,T)$ characterizes the DSW near the gradient catastrophe point both for the KdV equation with general initial data and also for a wide class of other dispersive equations, most of which are \emph{not} integrable.  Claeys and Grava \cite{ClaeysG09} of course proved Dubrovin's conjecture in the special case of the KdV equation with the help of the full power of complete integrability.  There is numerical evidence \cite{DubrovinGK11} that the conjecture holds for other non-integrable equations as well, but so far there is no proof.  A proof would likely be based on DZ-type analysis for a RH problem arising from perturbation theory to deform the KdV equation in the transform domain (breaking the simple time dependence of the reflection coefficient).  

More recently, Dubrovin, Grava, and Klein \cite{DubrovinGK09} carried out a similar formal analysis for dispersive perturbations of elliptic hydrodynamic problems (such as arise at the genus $N=0$ level from the focusing NLS equation). The type of singularity that appears generically at some point in the unperturbed problem with analytic initial data is called an \emph{elliptic umbilic catastrophe}.
The claim made in \cite{DubrovinGK09} is that in a neighborhood of the catastrophe point $(x_c,t_c)$ 
the dispersive problem has solutions that can be written in terms of the famous \emph{tritronqu\'ee} solution $u(\zeta)$ of the Painlev\'e-I equation, provided that when the suitably rescaled independent variables $(X,T)$ are combined into a single complex variable $\zeta$, the points $\zeta$ at which $u(\zeta)$ has (double) poles are avoided.  Near such points the dispersive terms are again important and determine the way the poles are regularized.  The study of Bertola and Tovbis \cite{BertolaT13} very nearly proves this conjecture in the case of the focusing NLS equation, with the only caveat being that, as is typical in such studies, the initial data is first replaced with an uncontrolled approximation that is valid in the semiclassical limit $\eps\to 0$ at $t=0$ but for which no estimates are directly available for nonzero $t$.  Remarkably, Bertola and Tovbis found that near the points in the $(x,t)$-plane corresponding to the poles of $u(\zeta)$, the solution of the focusing NLS equation can be approximated by a simple rescaling of the famous Peregrine breather solution \cite{Peregrine83} that is also important in the theory of rogue waves. 

\section{Ongoing work}
\subsection{Nonlocal problems}
There are a number of integrable dispersive equations in $1+1$ dimensions that are nonlocal in that they 
include terms involving integral transforms of the unknown wave field.  Examples include the Camassa-Holm equation, the intermediate long wave equation, and the Benjamin-Ono (BO) equation involving the Hilbert transform $\mathcal{H}$:
\begin{equation}
u_t + 2uu_x + \eps\mathcal{H}[u_{xx}]=0,\quad \mathcal{H}[f](x):=\frac{1}{\pi}\dashint_\mathbb{R}\frac{f(y)\,dy}{y-x},
\label{eq:BO}
\end{equation}
It is again of some interest to study the initial-value problem with initial data $u_0$ in the small-dispersion limit $\eps\to 0$.  On the formal side, this equation has multiphase wave solutions that are all rational functions of simple trigonometric phase factors, and the Whitham modulation equations for these $N$-phase waves take the form of $2N+1$ uncoupled copies of IB equation:  $u_{j,t} + 2u_ju_{j,x}=0$, $j=1,\dots,2N+1$ \cite{DobrokhotovK91}.  To study the generation of DSWs in this system,
formal analysis along the lines of the Gurevich-Pitaevskii \cite{GurevichP73} approach to KdV (cf., \S\ref{sec:GP}) was carried out independently by Jorge, Minzoni, and Smyth \cite{JorgeMS99} and Matsuno \cite{Matsuno98a,Matsuno98b}.
Appealing more to the complete integrability \cite{FokasA83,KaupM98}, an analogue of the LL method was developed for this problem in \cite{MillerX11}.  To date there is no fully-developed DZ method for the BO equation,
with the main obstruction being that the underlying RH problem is itself nonlocal.  However there is
some significant interest in developing such a method, because Masoero, Raimondo, and Antunes \cite{MasoeroRA15} have recently extended Dubrovin's perturbation analysis \cite{Dubrovin06} to a wider class of equations that includes BO, and they conjectured that for such systems the correct analogue of the function $U(X,T)$ from Theorem~\ref{thm:ClaeysG09} is a particular solution of a nonlocal analogue of the second Painlev\'e-I equation.  It would be very interesting to have a proof of this conjecture for the BO equation.
\subsection{Boundary-value problems}
Dispersive nonlinear equations can be formulated on domains smaller than the whole space $\mathbb{R}$, by including appropriate boundary conditions.  Such mixed initial-boundary value problems are frequently of interest in applications, for example, in studies of waves generated by a paddle in a wave tank.  Naturally, DSWs can be generated from forcing the system at the boundary as well as from initial data.
To study such problems from the point of view of complete integrability, a natural approach is to use
the so-called \emph{unified transform method} \cite{Fokas08}, which provides a relatively simple inverse method based on a RH problem, but for which the necessary scattering data is constrained by a nonlinear global relation that is difficult to unravel.  Asymptotic analysis for such integrable initial-boundary problems has been rather successful in the long-time and small-amplitude limits, but small-dispersion/semiclassical limits seem more challenging.  A recent result in the latter direction for the defocusing NLS equation, \eqref{eq:NLS-general} with $\sigma=1$, can be found in \cite{MillerQ15}, but the topic is basically wide open for further development.
\subsection{More independent variables}
One may also consider small-dispersion/semiclassical limits for integrable dispersive equations in more than one space dimension.  An example is to consider the semiclassical limit for the defocusing Davey-Stewartson II (dDSII) equations:
\begin{equation}
\begin{split}
i\epsilon \psi_t +\tfrac{1}{2}\epsilon^2\left(\psi_{xx}-\psi_{yy}\right) + 2M\psi&=0\\
M_{xx}+M_{yy}&=\left(|\psi|^2\right)_{yy}-\left(|\psi|^2\right)_{xx},
\end{split}
\label{eq:DSII}
\end{equation}
where $\psi=\psi(x,y,t)$ is complex and $M=M(x,y,t)$ is real.
The Cauchy problem for the dDSII system is to provide initial data $\psi(x,y,0)$ with suitable decay and seek a solution $\psi(x,y,t)$ for $t>0$ that also decays for large $r=\sqrt{x^2+y^2}$.  Under the assumption that $(|\psi|^2)_x$ and $(|\psi|^2)_y$ are $o(1/r)$ for large $r$, the auxiliary field $M$ is understood to be obtained for each $t\ge 0$ from $\psi$ via an integral against the free-space Green's function $\ln(r)$, and hence $M\to 0$ as $r\to\infty$.  The global well-posedness of this Cauchy problem has been established in the weighted Sobolev space $H^{1,1}(\mathbb{R}^2)$ by Perry \cite{Perry12}.  

To properly formulate 
the semiclassical limit,
we write $\psi$ in phase/amplitude form $\psi=Ae^{iS/\epsilon}$ and suppose that when $t=0$, both $A$ and $S$ are independent of $\epsilon$.  Introducing  Madulung's variables $\mathbf{u}:=\nabla S$ and $\rho:=A^2$, the system \eqref{eq:DSII} becomes
\begin{equation}
\begin{split}
\mathbf{u}_t +\frac{1}{2}\nabla \left(\mathbf{u}\cdot\sigma_3\mathbf{u}\right) -2\nabla M&=-\frac{\epsilon^2}{2}\nabla\left[\frac{\mathrm{div}\left(\sigma_3\nabla\sqrt{\rho}\right)}{\sqrt{\rho}}\right]\\
\rho_t + \mathrm{div}\left(\rho\sigma_3\nabla\mathbf{u}\right)&=0\\
\Delta M+\mathrm{div}\left(\sigma_3\nabla\rho\right)&=0.
\end{split}
\label{eq:MadelungDSII}
\end{equation}
In this form, it appears attractive to neglect the formally small term on the right-hand side, yielding the \emph{dispersionless dDSII} system.  The latter can only provide a good approximation of the true dynamics for small $\epsilon$ for sufficiently small time $t$ independent of $\epsilon$, because shocks can form \cite{KleinR14,ManakovS11}, after which time the neglected dispersive term cannot be ignored. 

The main obstruction to rigorous analysis of the semiclassical limit for the dDSII problem is actually the asymptotic approximation of the direct ST, which in this case involves the study of a linear elliptic $\overline{\partial}$-problem in the complex plane \cite{AblowitzF83,AblowitzF84,BealsC89,Fokas83,Sung94}.  One needs a suitable WKB-type method, not yet developed, to analyze this problem for $\eps\ll 1$.

\subsection{More dependent variables}
Finally, one may also consider integrable dispersive wave equations involving several coupled fields.
A canonical example is the three-wave resonant interaction system
\begin{equation}
\eps\psi_{j}^*\left(\psi_{j,t} + c_j\psi_{j,x}\right)=\gamma_j\psi_1^*\psi_2^*\psi_3^*,\quad j=1,2,3,
\end{equation}
where $\gamma_j=\pm 1$ and $c_j$ are distinct wave speeds.
This system is integrable by means of a Lax pair consisting of first-order $3\times 3$ linear systems, 
and leads to an inverse RH problem involving $3\times 3$ matrices \cite{Kaup76,ZakharovM75}.  An analysis of the direct spectral problem in the semiclassical limit is underway \cite{BuckinghamJM15}, but one
can see that for a DZ-type analysis of the inverse problem one needs a generalization of the $g$-function mechanism.    Such objects have been used in the study of so-called multiple orthogonal polynomials \cite{VanAsscheGK01} and in this situation the analogue of the maximization problem is played by a more challenging \emph{vector equilibrium problem}.

%
\section*{Acknowledgments}
The author was supported by the National Science Foundation under grant DMS-1513054.

%
%


\begin{thebibliography}{10}
\expandafter\ifx\csname url\endcsname\relax
  \def\url#1{\texttt{#1}}\fi
\expandafter\ifx\csname urlprefix\endcsname\relax\def\urlprefix{URL }\fi
\expandafter\ifx\csname href\endcsname\relax
  \def\href#1#2{#2} \def\path#1{#1}\fi
  
\bibitem{AblowitzF83}
M.~J.~Ablowitz and A.~S.~Fokas, ``Method of solution for a class of multidimensional nonlinear evolution equations,'' \textit{Phys.\@ Rev.\@ Lett.\@} \textbf{51}, 7--10, 1983.

\bibitem{AblowitzF84} 
M.~J.~Ablowitz and A.~S.~Fokas, ``On the inverse scattering transform of multidimensional nonlinear equations related to first-order systems in the plane,'' \textit{J.\@ Math.\@ Phys.\@} \textbf{25}, 2494--2505, 1984.

  
\bibitem{BealsC84}
R.~Beals and R.~R.~Coifman, ``Scattering and inverse scattering for first order systems,'' \textit{Comm.\@ Pure Appl.\@ Math.\@} \textbf{37}, 39--90, 1984.

\bibitem{BealsC89}
R.~Beals and R.~R.~Coifman, ``Linear spectral problems, nonlinear equations and the $\overline{\partial}$-method,'' \textit{Inv.\@ Prob.\@} \textbf{5}, 87--130, 1989.

\bibitem{BertolaT10}
M.~Bertola and A.~Tovbis,
``Universality in the profile of the semiclassical limit solutions to the focusing nonlinear Schr\"odinger equation at the first breaking curve,'' 
\textit{Int.\@ Math.\@ Res.\@ Not.\@ IMRN} \textbf{2010}, 2119--2167, 2010. 

\bibitem{BertolaT13}
M.~Bertola and A.~Tovbis,
``Universality for the focusing nonlinear Schr\"odinger equation at the gradient catastrophe point: rational breathers and poles of the tritronqu\'ee solution to Painlev\'e I,''
\textit{Comm.\@ Pure Appl.\@ Math.\@} \textbf{66}, 678--752, 2013. 

\bibitem{BlochK92}
A.~M.~Bloch and Y.~Kodama, 
``Dispersive regularization of the Whitham equation for the Toda lattice,'' 
\textit{SIAM J.\@ Appl.\@ Math.\@} \textbf{52}, 909--928, 1992. 

\bibitem{BuckinghamJM15}
R.~J.~Buckingham, R.~Jenkins, and P.~D.~Miller,
``On the three-wave interaction system in the semiclassical limit,''
in preparation, 2015.

\bibitem{BuckinghamM12}
R.~J.~Buckingham and P.~D.~Miller, 
``The sine-Gordon equation in the semiclassical limit: critical behavior near a separatrix,''
\textit{J.\@ Anal\@. Math\@} \textbf{118}, 397--492, 2012. Corrigenda in
\textit{J.\@ Anal.\@ Math.\@} \textbf{119}, 403--405, 2013. 

\bibitem{BuckinghamM13}
R.~J.~Buckingham and P.~D.~Miller, 
``The sine-Gordon equation in the semiclassical limit: dynamics of fluxon condensates,''
\textit{Mem.\@ Amer.\@ Math.\@ Soc.\@} \textbf{225}, 136 pp., 2013.

\bibitem{BuckinghamM14}
R.~J.~Buckingham and P.~D.~Miller, ``Large-degree asymptotics of rational Painlev\'e-II functions:  noncritical behaviour,'' \textit{Nonlinearity} \textbf{27}, 2489--2577, 2014.

\bibitem{ChesterFU57}
C.~Chester, B.~Friedman, and F.~Ursell,
``An extension of the method of steepest descents,''
\textit{Proc.\@ Cambridge Philos.\@ Soc.\@} \textbf{53}, 599--611, 1957. 

\bibitem{ClaeysG09}
T.~Claeys and T.~Grava, ``Universality of the break-up profile for the {K}d{V}
  equation in the small dispersion limit using the {R}iemann-{H}ilbert
  approach,'' \textit{Comm.\@ Math.\@ Phys.\@} \textbf{286}, 979--1009, 2009.
  
\bibitem{ClaeysG10a}
T.~Claeys and T.~Grava, ``Solitonic asymptotics for the Korteweg-de Vries equation in the small dispersion limit,'' \textit{SIAM J.\@ Math.\@ Anal.\@} \textbf{42}, 2132--2154, 2010.

\bibitem{ClaeysG10b}
T.~Claeys and T.~Grava, ``Painlev\'e II asymptotics near the leading edge of the oscillatory zone for the Korteweg-de Vries equation in the small-dispersion limit,'' \textit{Comm.\@ Pure Appl.\@ Math.\@} \textbf{63}, 203--232, 2010. 
  
\bibitem{DeiftKMVZ99a}
P.~Deift, T.~Kriecherbauer, K.~T.-R.~McLaughlin, S.~Venakides, and X.~Zhou,
``Uniform asymptotics for polynomials orthogonal with respect to varying exponential weights and applications to universality questions in random matrix theory,'' 
\textit{Comm.\@ Pure Appl.\@ Math.\@} \textbf{52}, 1335--1425, 1999. 

\bibitem{DeiftKMVZ99b}
P.~Deift, T.~Kriecherbauer, K.~T.-R.~McLaughlin, S.~Venakides, and X.~Zhou,
``Strong asymptotics of orthogonal polynomials with respect to exponential weights,''
\textit{Comm.\@ Pure Appl.\@ Math.\@} \textbf{52} 1491--1552, 1999. 


\bibitem{DeiftVZ94}
P.~Deift, S.~Venakides, and X.~Zhou,
``The collisionless shock region for the long-time behavior of solutions of the KdV equation,'' \textit{Comm.\@ Pure Appl.\@ Math.\@} \textbf{47}, 199--206, 1994. 

\bibitem{DeiftVZ97}
P.~Deift, S.~Venakides, and X.~Zhou,
``New results in small dispersion KdV by an extension of the steepest descent method for Riemann-Hilbert problems,'' \textit{IMRN Internat.\@ Math.\@ Res.\@ Notices} \textbf{1997}, 285--299, 1997.

\bibitem{DeiftZ92}
P.~Deift and X.~Zhou,
``A steepest descent method for oscillatory Riemann-Hilbert problems,''
\textit{Bull.\@ Amer.\@ Math.\@ Soc.\@} \textbf{26} 119--123, 1992.

\bibitem{DeiftZ93}
P.~Deift and X.~Zhou,
``A steepest descent method for oscillatory Riemann-Hilbert problems. Asymptotics for the MKdV equation,''
\textit{Ann.\@ Math.\@} \textbf{137}, 295--368, 1993. 

\bibitem{DifrancoM13}
J.~C.~DiFranco and P.~D.~Miller, ``The semiclassical modified nonlinear Schr\"odinger equation II:  Asymptotic analysis of the Cauchy problem.  The elliptic region for transsonic initial data,'' \textit{Contemp.\@ Math.\@} \textbf{593}, 29--81, 2013.

\bibitem{DifrancoMM11}
J.~C.~DiFranco, P.~D.~Miller, and B.~K.~Muite, ``On the modified nonlinear Schr\"odinger equation in the semiclassical limit:  supersonic, subsonic, and transsonic behavior,'' \textit{Acta Math.\@ Sci.\@} \textbf{31B}, 2343--2377, 2011.
 
\bibitem{DobrokhotovK91}
S.~Yu.~Dobrokhotov and I.~M.~Krichever, ``Multi-phase solutions of the {B}enjamin-{O}no equation and their averaging,'' \textit{Mat.\@ Zametki} \textbf{49}, 42--58, 1991; English translation in \textit{Math.\@ Notes} \textbf{49}, 583--594, 1991.

\bibitem{Dubrovin06}
B.~Dubrovin, ``On Hamiltonian perturbations of hyperbolic systems of conservation
  laws, {II}: {U}niversality of critical behaviour,'' \textit{Comm.\@ 
  Math.\@ Phys.\@} \textbf{267}, 117--139, 2006.

\bibitem{DubrovinGK09}
B.~Dubrovin, T.~Grava, and C.~Klein, 
``On universality of critical behavior in the focusing nonlinear Schr\"odinger equation, elliptic umbilic catastrophe and the tritronqu\'ee solution to the Painlev\'e-I equation,''
\textit{J.\@ Nonlinear Sci.\@} \textbf{19}, 57--94, 2009. 

\bibitem{DubrovinGK11} B.~Dubrovin, T.~Grava, and C.~Klein, ``Numerical study of breakup in generalized Korteweg-de Vries and Kawahara equations,'' \textit{SIAM J.\@ Appl.\@ Math.\@} \textbf{71}, 983--1008, 2011.

  
\bibitem{FlaschkaFM80}
H.~Flaschka, M.~G.~Forest, and D.~W.~McLaughlin, 
``Multiphase averaging and the inverse spectral solution of the Korteweg-de Vries equation,''
\textit{Comm.\@ Pure Appl.\@ Math.\@} \textbf{33}, 739--784, 1980. 

\bibitem{Fokas83} A~ S.~Fokas, ``Inverse scattering of first-order systems in the plane related to nonlinear multidimensional equations,'' \textit{Phys.\@ Rev.\@ Lett.\@} \textbf{51}, 3--6, 1983.

\bibitem{Fokas08}
A.~S.~Fokas, \textit{A Unified Approach to Boundary Value Problems,} CBMS-NSF Regional Conference Series in Applied Mathematics, Society for Industrial and Applied Mathematics, Philadelphia, 2008.

\bibitem{FokasA83}
A.~S.~Fokas and M.~J.~Ablowitz,
``The inverse scattering transform for the Benjamin-Ono equation --- a pivot to multidimensional problems,''  \textit{Stud.\@ Appl.\@ Math.\@} \textbf{68}, 1--10, 1983.

\bibitem{Gerard93}
P.~G\'erard,
``Remarques sur l'analyse semi-classique de l'\'equation de Schr\"odinger non lin\'eaire'' (French) [Remarks on the semiclassical analysis of the nonlinear Schr\"odinger equation], \textit{S\'eminaire sur les \'Equations aux D\'eriv\'ees Partielles, 1992--1993} \textbf{13}, 13 pp., \'Ecole Polytech.\@, Palaiseau, 1993. 

\bibitem{Grenier95}
E.~Grenier, 
``Limite semi-classique de l'\'equation de Schr\"odinger non lin\'eaire en temps petit. (French. English, French summary) [Semiclassical limit of the nonlinear Schr\"odinger equation in small time] 
\textit{C.\@ R.\@ Acad.\@ Sci.\@ Paris S\'er.\@ I Math.\@} \textbf{320}, 691--694, 1995. 


\bibitem{GurevichP73}
A.~V.~Gurevich and L.~P.~Pitaevskii, ``Nonstationary structure of a collisionless shock wave,'' \textit{Zh.\@ \`Eksper.\@ Teoret.\@ Fiz.\@} \textbf{65}, 590--604, 1973; (Russian) English translation in \textit{Sov.\@ Phys.\@ JETP} \textbf{38}, 291--297, 1974.
  
\bibitem{JinLM99}
Shan Jin, C.~D.~Levermore, and D.~W.~McLaughlin, ``The semiclassical limit of the defocusing NLS hierarchy,'' \textit{Comm.\@ Pure Appl.\@ Math.\@} \textbf{52}, 613--654, 1999.

\bibitem{JorgeMS99}
M.~C.~Jorge, A.~A.~Minzoni, and N.~F.~Smyth, 
``Modulation solutions for the Benjamin-Ono equation,''
\textit{Phys.\@ D} \textbf{132}, 1--18, 1999.

\bibitem{KamvissisMM03}
S.~Kamvissis, K.~D.~T.-R.~McLaughlin, and P.~D.~Miller, 
\textit{Semiclassical soliton ensembles for the focusing nonlinear Schršdinger equation,}
Annals of Mathematics Studies, \textbf{154}, Princeton University Press, Princeton, NJ, 2003.


\bibitem{Kaup76}
D.~J.~Kaup, ``The three-wave interaction --- a nondispersive phenomenon,'' \textit{Stud.\@ Appl.\@ Math.\@} \textbf{55}, 9--44, 1976.

\bibitem{KaupM98}
D.~J.~Kaup and Y.~Matsuno,
``The inverse scattering transform for the Benjamin-Ono equation,''
\textit{Stud.\@ Appl.\@ Math.\@} \textbf{101}, 73--98, 1998.

\bibitem{KleinR14}
C.~Klein and K.~Roidot, 
``Numerical study of the semiclassical limit of the Davey-Stewartson II equations,''
\textit{Nonlinearity} \textbf{27}, 2177--2214, 2014. 


\bibitem{LaxL83}
P.~D.~Lax and C.~D.~Levermore, ``The small dispersion limit of the {K}orteweg-de
  {V}ries equation. {I}, {II}, {III},'' \textit{Comm.\@ Pure Appl.\@
  Math.\@} \textbf{36}, 253--290, 571--593, 809--829, 1983.
  
\bibitem{Madelung26}
E.~Madelung, ``Quantum theory in hydrodynamic form,'' 
\textit{Zeitschr.\@ Phys.\@} \textbf{40}, 322--326, 1926.

\bibitem{ManakovS11} S.~V.~Manakov and P.~M.~Santini, ``Solvable vector nonlinear Riemann problems, exact implicit solutions of dispersion less PDEs and wave breaking,'' \textit{J.\@ Phys.\@ A} \textbf{44}, article number 345203, 2011.


\bibitem{MasoeroRA15}
D.~Masoero, A.~Raimondo, and P.~R.~Antunes, ``Critical behavior for scalar nonlinear
  waves,'' \textit{Phys.\@ D} \textbf{292--293}, 1--7, 2015.

\bibitem{Matsuno98a}
Y.~Matsuno, ``The small dispersion limit of the {B}enjamin-{O}no equation and the
  evolution of a step initial condition,'' \textit{J. Phys.\@ Soc.\@ Japan} \textbf{67}, 1814--1817, 1998.

\bibitem{Matsuno98b}
Y.~Matsuno, ``Nonlinear modulation of periodic waves in the small dispersion limit of the Benjamin-Ono equation,'' \textit{Phys.\@ Rev.\@ E} \textbf{58}, 7934--7940, 1998.
  
\bibitem{McIntoshCM82}
R.~R.~Coifman, A.~McIntosh, and Y.~Meyer, 
``L'int\'egrale de Cauchy d\'efinit un op\'erateur born\'e sur $L^2$ pour les courbes lipschitziennes,'' (French) [The Cauchy integral defines a bounded operator on $L^2$ for Lipschitz curves],
\textit{Ann.\@ Math.\@} \textbf{116}, 361--387, 1982. 

\bibitem{Miller06}
P.~D.~Miller, 
\textit{Applied Asymptotic Analysis,} 
Graduate Studies in Mathematics, \textbf{75}, American Mathematical Society, Providence, RI, 2006. 

\bibitem{Miller08}
P.~D.~Miller,
``Riemann-Hilbert problems with lots of discrete spectrum,'' \textit{Contemp.\@ Math.\@} \textbf{458}, 163--181, 2008.

\bibitem{MillerQ15}
P.~D.~Miller and Z.~Y.~Qin, ``Initial-boundary value problems for the defocusing nonlinear Schr\"odinger equation in the semiclassical limit,'' \textit{Stud.\@ Appl.\@ Math.\@} \textbf{134}, 276--362, 2015.

\bibitem{MillerX11}
P.~D.~Miller and Z.~Xu, ``On the zero-dispersion limit of the {B}enjamin-{O}no
  {C}auchy problem for positive initial data,'' \textit{Comm.\@ Pure Appl.\@ Math.\@} \textbf{64}, 205--270, 2011.
  
\bibitem{dlmf}
NIST Digital Library of Mathematical Functions. \\
\texttt{http://dlmf.nist.gov/}, Release 1.0.10 of 2015-08-07. 
Online companion to \cite{OlverLBC10}.

\bibitem{OlverLBC10}
F.~W.~J.~Olver, D.~W.~Lozier, R.~F.~Boisvert, and C.~W.~Clark, editors,
\textit{NIST Handbook of Mathematical Functions,}
Cambridge University Press, New York, NY, 2010. 
Print companion to \cite{dlmf}.

\bibitem{Peregrine83} D.~H.~Peregrine, ``Water waves, nonlinear Schr\"odinger equations and their solutions,'' \textit{J.\@ Austral.\@ Math.\@ Soc.\@ Ser.\@ B} \textbf{25}, 16--43, 1983.

\bibitem{Perry12}
P.~A.~Perry, ``Global well-posedness and long-time asymptotics for the defocussing Davey-Stewartson II equation in $H^{1,1}(\mathbb{R}^2)$,'' \texttt{arXiv:1110.5589v2}, 2012.

\bibitem{Ramond96}
T.~Ramond, 
``Semiclassical study of quantum scattering on the line,''
\textit{Comm.\@ Math.\@ Phys.\@} \textbf{177}, 221--254, 1996. 

\bibitem{Sung94} L.-Y.~Sung, ``An inverse scattering transform for the Davey-Stewartson II equations.  I, II, III,'' \textit{J.\@ Math.\@ Anal.\@ Appl.\@} \textbf{183}, 121--154, 289--325, 477--494, 1994.

\bibitem{Szego75}
G.~Szeg\H{o},
\textit{Orthogonal polynomials,} 
Fourth edition. American Mathematical Society, Colloquium Publications, Vol. XXIII. American Mathematical Society, Providence, R.I., 1975. 

\bibitem{TovbisVZ04}
A.~Tovbis, S.~Venakides, and X.~Zhou,
``On semiclassical (zero dispersion limit) solutions of the focusing nonlinear Schršdinger equation,''
\textit{Comm.\@ Pure Appl.\@ Math.\@} \textbf{57}, 877--985, 2004. 


\bibitem{TracyW94}
C.~A.~Tracy and H.~Widom,
``Level-spacing distributions and the Airy kernel,''
\textit{Comm.\@ Math.\@ Phys.\@} \textbf{159}, 151--174, 1994. 

\bibitem{Tsarev85}
S.~P.~Tsar\"ev, 
``Poisson brackets and one-dimensional Hamiltonian systems of hydrodynamic type,'' (Russian)
\textit{Dokl.\@ Akad.\@ Nauk SSSR} \textbf{282}, 534--537, 1985. 

\bibitem{VanAsscheGK01} W.~Van Assche, J.~Geronimo, and A.~Kuijlaars,
  ``Riemann-Hilbert problems for multiple orthogonal polynomials,'' in
  \textit{Special functions 2000: current perspective and future
    directions (Tempe, AZ)}, 23--59, \textit{NATO Sci. Ser. II
    Math. Phys. Chem.}, \textbf{30}, Kluwer Acad. Publ., Dordrecht,
  2001.


\bibitem{Venakides90}
S.~Venakides, 
``The Korteweg-de Vries equation with small dispersion: higher order Lax-Levermore theory,'' 
\textit{Comm.\@ Pure Appl.\@ Math.\@} \textbf{43}, 335--361, 1990. 

\bibitem{Whitham99}
G.~B.~Whitham, 
\textit{Linear and nonlinear waves,}
Reprint of the 1974 original. Pure and Applied Mathematics (New York). A Wiley-Interscience Publication. John Wiley \& Sons, Inc., New York, 1999. 

\bibitem{ZakharovM75} V.~E.~Zakharov and S.~V.~Manakov, 
``The theory of resonance interaction of wave packets in nonlinear media,''
\textit{Sov. Phys. JETP} \textbf{42}, 842--850, 1975.


\bibitem{ZakharovS73}
V.~E.~Zakharov and A.~B.~Shabat, ``Interaction between solitons in a stable medium,'' \textit{Sov. Phys. JETP} \textbf{37}, 823--828, 1973.

\end{thebibliography}
\end{document}